\documentclass[draft]{agujournal2019}
\usepackage{url} 
\usepackage{soul}
\graphicspath{{figures/}}
\draftfalse

\journalname{ JGR: Earth Surface}

\begin{document}

%
%


\title{Dune initiation in a bimodal wind regime}

%
%

\authors{Pauline Delorme\affil{1}, Giles F.S. Wiggs\affil{2}, Matthew C. Baddock\affil{3}, Philippe Claudin\affil{4}, Joanna M. Nield\affil{1}, and Andrew Valdez\affil{5}}

\affiliation{1}{School of Geography and Environmental Science, University of Southampton, UK}

\affiliation{2}{School of Geography and the Environment, University of Oxford, UK}

\affiliation{3}{Geography and Environment, Loughborough University, UK}

\affiliation{4}{Laboratoire de Physique et Mécanique des Milieux hétérogènes, PMMH UMR7636 CNRS, ESPCI Paris, PSL University, Sorbonne Université, Université de Paris, France}

\affiliation{5}{National Park Service – NPS-Great Sand Dunes National Park $\&$ Preserve, USA}

\correspondingauthor{Pauline Delorme}{P.M.T.Delorme@soton.ac.uk}

\vspace{3cm}
\textcolor{blue}{An edited version of this paper was published by AGU. Copyright 2020 American Geophysical Union: Delorme, P., Wiggs, G. F. S., Baddock, M. C., Claudin, P., Nield, J. M., $\&$ Valdez, A. (2020). Dune initiation in a bimodal wind regime. \textit{Journal of Geophysical Research: Earth Surface}, 125, e2020JF005757. https://doi.org/10.1029/2020JF005757}

\begin{keypoints}
\item We report high-resolution measurements of protodune morphodynamics during a one-month field campaign.
\item Our data provide the first field validation of a linear stability analysis for a flat sand bed under a bi-directional wind regime.
\item We quantify relations between sand flux and complex wind regime characteristics that are important for aeolian dune initiation and growth.
\end{keypoints}


\begin{abstract}
Early-stage bedforms develop into mature dunes through complex interactions between wind, sand transport and surface topography. Depending on varying environmental and wind conditions, the mechanisms driving dune formation and, ultimately, the shape of nascent dunes may differ markedly. In cases where sand availability is plentiful, the emergence and growth of dunes can be studied with a linear stability analysis of coupled transport and hydrodynamic equations. Until now, this analysis has only been applied using field evidence in uni-directional winds. However, in many areas of the world and on other planets, wind regimes are more often bimodal or multimodal. Here, we investigate field evidence of protodune formation under a bimodal wind regime by applying linear stability analysis to a developing protodune field. Employing recent development of the linear stability theory and experimental research, combined with in-situ wind, sediment transport, and topographic measurements during a month-long field campaign at Great Sand Dunes National Park, Colorado, USA, we predict the spatial characteristics (orientation and wavelength) and temporal evolution (growth rate and migration velocity) of a protodune field. We find that the theoretical predictions compare well with measured dunefield attributes as characterized by high-resolution Digital Elevation Models measured using repeat terrestrial laser scanning. Our findings suggest that linear stability analysis is a quantitative predictor of protodune development on sandy surfaces with a bimodal wind regime. This result is significant as it offers critical validation of the linear stability analysis for explaining the initiation and development of dunes towards maturity in a complex natural environment.
\end{abstract}


 \section{Introduction}

The variability in sand dune patterns is, in part, due to complex interactions between fluid flow, bed morphology and sediment transport. The coupling between flow and sediment transport controls dune morphology and, concurrently, there is a strong modification of airflow dynamics driven by dune topography \cite{wiggs1996role, walker2002dynamics,wiggs2012turbulent}. Although these interactions are well studied on mature dunes, we lack field observations of early-stage (proto) dunes because the small scale of relevant morphological features and associated flow processes are extremely difficult to measure \cite{kocurek2010bedform}. 

The early stages of dune development involve localized sand deposition, which, under conditions of positive flow-transport-form feedback, leads to dune emergence. To understand the mechanisms of initial dune nucleation and subsequent growth it is therefore necessary to quantify the relationships between flow, sediment transport and morphology close to the surface. Although requisite process measurements are challenging on small bedforms, the quantification of these fundamental relationships on protodunes is crucial because of the importance this bedform type has as a precursor to fully developed dunes \cite{kocurek2010bedform,claudin2013field,baddock2018early,phillips2019low}. 
 
At this small scale, a range of theoretical studies, laboratory experiments and physically-based models have established quantitative predictions of fluid dynamic and sand transport controls on bedform development for both subaqueous and aeolian environments \cite{richards1980formation,sauermann2001continuum,andreotti2002selection,andreotti2010measurements, fourriere2010bedforms, duran2011aeolian}. In recent decades, continuous dune models have been developed to describe the shape of barchan and transverse dunes in uni-directional wind regimes \cite{andreotti2002selection,parteli2006profile}, and cellular automaton models have been used to study protodune \cite{nield2011surface} and dune dynamics \cite <e.g.>[]{narteau2009setting,zhang2010morphodynamics,eastwood2011modelling,worman2013modeling,rozier2014real}. In addition, the emergence of dunes on an erodible sand bed has been investigated theoretically through linear stability analysis \cite{kennedy1963mechanics,richards1980formation, andreotti2002selection,colombini2004revisiting,kouakou2005stability,claudin2006scaling,devauchelle2010stability, andreotti2012bedforms}. Linear stability analysis shows that small perturbations in surface morphology are amplified by an instability process that results from the balance between a destabilizing hydrodynamical mechanism and a countering stabilising sediment transport mechanism. This leads eventually to bedform emergence, as discussed by \citeA{fourriere2010bedforms}, \citeA{duran2011aeolian}, \citeA{ping2014emergence}, \citeA{courrech2014two} and, \citeA{gadalspatial}. Independant field measurements of the flow and sediment transport over a small bed perturbation have previously quantified the two mechanisms controlling the instability process \cite{elbelrhiti2005field,andreotti2010measurements,claudin2013field}. Only recently, however, \citeA{gadalspatial} have shown the validity of the linear stability analysis to predict the emergence and growth of incipient dunes in natural aeolian environments characterised by uni-directional wind regimes.

Despite the often high variability of the transport-competent winds in the natural environment, only a few studies have focused attention on dune dynamics in multi-directional wind regimes \cite{rubin1987bedform,rubin1990flume, gao2015phase}. These models show that under such environmental conditions, dunes tend to align to maximize the gross sand transport occurring normal to the crest \cite{rubin1987bedform}. In the case of a bimodal wind regime, water tank experiments demonstrated that the crucial parameter controlling the dune orientation is the divergent angle \cite <i.e. the angle between the two winds, >[]{parteli2009dune,reffet2010formation, courrech2014two}. \citeA{parteli2009dune} and \citeA{reffet2010formation} observed that transverse dunes formed with divergent angles smaller than 90$^{\circ}$, whereas longidunal dunes were observed when the divergent angle is greater than 90$^{\circ}$. Whilst the dependence of dune orientation and pattern on wind regime has been validated only rarely in natural aeolian environments \cite <e.g.>[]{ping2014emergence,courrech2014two,gao2018morphodynamics}, the linear stability analysis has recently been extended theoretically to multi-directional winds \cite{gadal2019incipient}.   

In this paper, we undertake the first investigation of the emergence of protodunes in a natural aeolian sandy environment through linear stability analysis in a bimodal wind regime. Our field site was located in the Great Sand Dunes National Park (Colorado, USA) and consisted of a flat sandy surface that facilitated the emergence of protodunes. During a one-month field campaign we measured the wind and sediment transport characteristics at the site and monitored the evolution of the growing bedforms through repeat terrestrial laser scanning (TLS). Using an adaptation of the linear stability analysis incorporating wind bimodality, we examined the measured wind and sand transport data to recover predicted information for dune orientation ($\alpha$), wavelength ($\lambda$), growth rate ($\sigma$), and migration velocity ($c$). We compared these predicted data to those measured with the TLS to evaluate the explanatory power of linear stability analysis for the initiation of dunes in a bimodal wind regime.

\section{Methods}
 
\subsection{Field description}

\begin{figure*}[h!]
\centering
\includegraphics[width=\linewidth]{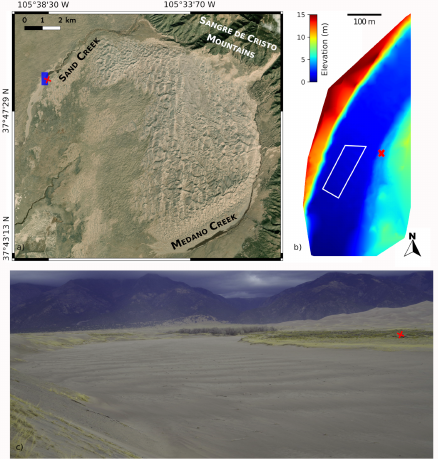}
\caption{Field characteristics. a) Satellite image of the Great Sand Dunes National Park dunefield. The blue rectangle indicates the field site in Sand Creek (Source: Esri). b) TLS-derived Digital Elevation Model of the Sand Creek study site. The white polygon shows the area used for the morphometric analysis of protodunes. c) Photograph of the Sand Creek study site, looking toward the North from the western edge of the creek. Protodunes (the lighter grey strips of finer sand, 0.2 m high) can be seen extending W-E across the bed of the ephemeral channel. The red cross in each figure indicates the location of the TLS during survey measurements.
\label{fig:localisation}}
\end{figure*}

The Great Sand Dunes National Park is located in the San Luis Valley, in south-central Colorado (USA). The prevailing southwesterly and westerly winds in the San Luis Valley transport sand toward the foot of the Sangre de Cristo mountains creating a complex dunefield system of transverse and barchanoid dunes \cite{madole2008origin}. The dunefield is bounded by two ephemeral channels: Medano Creek in the south-east and Sand Creek in the north-west (Fig. \ref{fig:localisation}a). Each year in late spring and early summer snow melt feeds the two creeks such that there is a fluvial redistribution of sediment, resetting the creek beds to a flat surface \cite{valdez1996role}. This phenomenon helps establish these creek sites as a novel setting to investigate drivers of dune initiation and development because winds act on these wide, flat, sandy surfaces resulting in the annual development of protodunes on the beds of the channels. The characteristics of these creek sites make them suitable for landscape-scale experiments that are particularly well suited for validation and quantification purposes \cite{wilcock2008need,ping2014emergence}. 
Our experimental field site was established at Sand Creek (Fig. \ref{fig:localisation}a). This is the more active of the two bounding channels and so the resetting (flattening) of its bed by spring floodwater tends to be more extensive \cite{valdez1996role}.

\subsection{Measuring wind and sediment transport characteristics}
Data were collected at the Sand Creek field site from March 23$^{\mathrm{rd}}$ to April 16$^{\mathrm{th}}$ 2019 (24 days). For the duration of the experiment a temporary weather station was erected adjacent to the TLS measurement site (Fig. \ref{fig:localisation}c). Here wind speed and direction were measured at a height of 3.17 m using a Gill 2D sonic anemometer with co-located measurements of saltation activity derived from a Sensit mass transport monitor \cite{gillette1986mass,stockton1990field}. The anemometer and Sensit were connected to a Campbell CR800 Series datalogger with data recorded at a frequency of 0.1 Hz. For each time step (10 s), we extracted mean wind speed $u_i$ and wind direction $\overrightarrow{x_i}$.  
Using the wind velocity and Sensit records, we defined the threshold wind velocity for sand transport, $u_{\mathrm{th}}$ by computing the maximum saltation counts occurring in each 0.01~m$\,$s$^{-1}$ bin of velocity data (Fig. \ref{fig:field_data}c). Employing the law-of-the-wall and assuming a logarithmic velocity increase with height above the surface, we also calculated the shear velocity, $u_*$ (analogous to basal shear stress);
 
\begin{equation}
 u_{*} = \frac{u(z) \kappa}{\mathrm{ln}\left(\frac{z}{z_0}\right)}, 
\label{eq:law_of_wall}
\end{equation}
where $u$ is the wind velocity at height $z$, $\kappa$ = 0.4 (the von-K\'arm\'an constant), and $z_0$ is the aerodynamic roughness length, due to the transport layer, generally of the order of a millimetre \cite{duran2011aeolian,claudin2013field, gadal2019incipient}. 
   
Wind tunnel experiments have highlighted empirical relations between sand flux and shear velocity, $u_*$ \cite{bagnold1937transport,ungar1987steady,iversen1999effect,duran2011aeolian}. For low to moderate wind speeds, when the shear velocity $u_*$ is higher than a threshold shear velocity $u_{*\mathrm{th}}$ the sediment flux, $Q_s$, evolves quadratically on a sandy erodible surface \cite{ungar1987steady,rasmussen1996saltation, andreotti2004two,creyssels2009saltating,ho2011scaling, pahtz2020unification,ralaiarisoa2020transition}, 

\begin{equation}
Q_s = 25 \frac{\rho}{\rho_s} \sqrt{\frac{d}{g}}  \left( u_*^2 - u_{*\mathrm{th}}^2 \right),
\label{eq:transport_law}
\end{equation}
where $d$ is the average grain diameter, $\rho$ and $\rho_s$ are respectively the air and sediment density, and $g$ is the gravitational acceleration. To determine $d$ and $\rho_s$, we collected surface samples of dune and inter-dune sand. The grain diameter was established using a laser granulometer (Malvern Mastersizer 3000 laser), and the density of the sand using a pycnometer.

To characterize the wind regime, we calculated the resultant drift potential (RDP), drift potential (DP), and the resultant drift direction (RDD), defined as the direction of the resultant sand flux \cite{fryberger1979dune, bullard1996wind}. 

\begin{equation}
DP = \frac{ \sum_{i= 2}^{N}\Vert \overrightarrow{Q_i} \Vert  \delta t_i}{\sum_{i= 2}^{N}\delta t_i},
\label{eq:DP}
\end{equation}

\begin{equation}
RDP = \frac{\Vert \sum_{i= 2}^{N} \overrightarrow{Q_i} \delta t_i \Vert}{\sum_{i= 2}^{N}\delta t_i},
\label{eq:RDP}
\end{equation}
where $N$ is the angular sector (i.e. the wind direction for each 1$^\circ$ bin ), and $Q_i$ is the associated individual flux vector. The ratio of $RDP/DP$ is a non-dimensional number used to characterize the directional variability of the wind, where values closer to unity represent a more uni-directional wind regime.

\subsection{Protodune morphology}

\subsubsection{Measuring dynamic topography}

Throughout the period of the experiment high-resolution scans of the surface topography, repeated every 3-7 days, were undertaken using a Leica P50 Scanstation TLS on a tripod at a height of 1.5 m and with a resolution setting of 0.0016 m at 10 m for an instrument sensitivity of 1,000 m. Five scans were performed, respectively, on the 27$^{\mathrm{th}}$ and 30$^{\mathrm{th}}$ of March, and 5$^{\mathrm{th}}$, 12$^{\mathrm{th}}$, and 16$^{\mathrm{th}}$ of April 2019. The placement of the TLS (see Fig. \ref{fig:localisation}) enabled measurement scans to span the entire width of the channel encompassing multiple bedforms along the channel long axis. The instrument placement also minimised both surface occlusion and the distance to the bedforms of interest (maximum distance of 200 m from the TLS). No human or animal disturbance of the surface was apparent during the measurement period. In order to measure the gradient of the channel bed in the absence of protodunes, a further TLS scan of the surface was carried out in November 2019 using a Leica C10 Scanstation TLS with a resolution setting of 0.001 m vertical and 0.005 m horizontal at 10 m. The surface measured in November 2019 had been reset by fluvial activity in the preceding summer which was extensive enough to inundate the full channel. In order to follow the temporal evolution of the developing protodunes, scans were aligned using five fixed targets established outside of the measurement area. This allowed us to achieve a mean absolute error of registered scans of 0.005 m across the entire March to November measurement period. A handheld GPS was used to orientate the registered surfaces to true north using the TLS location and a fixed target.

\subsubsection{TLS post-processing and surface detrending}

Analysis of the TLS data involved extracting a rectangular section of measured points (approximately 340 m long and 55 m wide) along the centreline of the channel to facilitate the determination of protodune morphology and migration rates and also to eliminate the effect of the channel banks on the area under investigation (Fig. \ref{fig:localisation}b). Each extracted rectangle contained approximately 3.2 million measured data points. Point data were filtered to separate surface and active saltation points following the methods of \citeA{nield2011application} giving a mean surface point density of 775 points/m$^2$. Surface points were gridded with a resolution of 0.1 m and interpolated using Matlab Natural Neighbour following the methods of \citeA{nield2013estimating}. Finally, surfaces were de-trended for the underlying channel gradient ($\sim$0.003) by removing the smoothed topographic surface surveyed in November 2019 using a 15 m by 15 m moving window, following the methods of \citeA{hugenholtz2010spatial}, thus producing five detrended Digital Elevation Model (DEM) surfaces throughout the spring measurement period.

\subsubsection{Calculation of the protodune morphometrics and migration rates}
 
The time series of DEMs resulting from the TLS surveys included six complete protodune forms. We analysed the DEMs, to measure the morphological and dynamic characteristics of the protodunes for comparison with those predicted by the linear stability analysis. The TLS-derived DEMs were inspected using a "peak-by-peak" analysis, to extract the spatial characteristics (wavelength and orientation) and temporal evolution (migration celerity and growth rate) of the protodune field. The "peak-by-peak" analysis consisted of extracting the peak positions (minimum and maximum) on seven transects perpendicular to the dune crest, with a regular across-channel spacing of 7 m between each transect. For each DEM, we used the position of the maximum peak of each protodune to calculate the orientation of each bedform and the distance between them. The displacement of the maximum peak between the first and last DEM allowed us to estimate the migration rate for each protodune. Finally we calculated the variation in amplitude of each protodune using the same peak-by-peak method to determine growth rates (details are provided in the Supporting Information).
 
\subsection{Mechanisms of dune instability}
 
The bed instability that results in dune formation is a consequence of the interaction between sand bed, flow and sediment transport \cite{devauchelle2010stability,andreotti2012bedforms}. Over a small topographic irregularity on a sandy surface, the simultaneous hydrodynamical effects of inertia and dissipation cause a phase advance of the surface shear stress \cite <i.e. the maximum shear stress is shifted upwind of the crest,>[]{jackson1975turbulent,claudin2006scaling,duran2011aeolian, claudin2013field}. This phase advance can be mathematically characterized by two hydrodynamical parameters $\cal{A}$, which is in-phase with the topography, and $\cal{B}$, which is in phase quadrature with the topography. This hydrodynamical mechanism represents a destabilizing process. In addition, the relaxation of the sediment flux toward an equilibrium value generates an additional spatial phase lag. This spatial lag is characterised by the saturation length, $L_{\mathrm{sat}}$. In practice, the saturation length scales as the distance needed for one grain to be accelerated to the velocity of the wind \cite{hersen2002relevant, claudin2006scaling, duran2011aeolian, charru2013sand, pahtz2013flux,selmani2018aeolian}. The prefactor of the scaling has been validated with wind tunnel experiments \cite{andreotti2010measurements},

\begin{equation}
L_{\mathrm{sat}} = 2.2\, \frac{\rho_s}{\rho} d.
\label{eq:L_sat}
\end{equation}

The effect of the saturation length is such that the maximum sediment flux is shifted downwind of the maximum shear stress, representing a stabilizing process. The hydrodynamical (destabilizing) effect and the associated relaxation of the sediment flux (stabilizing effect) are the two main contributors to the phase shift and instability between topography and sediment flux \cite{fourriere2010bedforms}. The crest of a protodune is therefore located in a deposition zone, a situation that induces its vertical growth \cite{kennedy1963mechanics,naqshband2017sharp}. The aim of the linear stability analysis is to determine the characteristics (wavelength, $\lambda$ and orientation, $\alpha$) of the most unstable (growth) mode generated by the simultaneous effects of the stabilizing and destabilizing mechanisms. This can be achieved by solving the fundamental dispersion relations of growth rate, $\sigma$ and propagation velocity, $c$ \cite{andreotti2012bedforms, charru2013sand}. 

In the case of a bimodal wind regime such as that observed at Sand Creek (Fig. \ref{fig:field_data}b and d) the growth rate ($\sigma_{\Sigma}$) and propagation velocity ($\overrightarrow{c}_{\Sigma}$) can be expressed as the sum of the contribution of each effective wind (i.e. the wind above threshold that is associated with sand transport). Following \citeA{gadal2019incipient}, the dimensional equations read,  

\begin{equation}
\sigma_{\Sigma}(\alpha, k) = \frac{1}{L^2_{sat}} \frac{\sum_i t_i \, \frac{Q_i u^2_{* i}}{u^2_{* i} - u^2_{th*}}\, \sigma(u_{* i}, \alpha_i, k)}{\sum_i t_i} ,
\label{eq:growth_sum}
\end{equation} 

\begin{equation}
\overrightarrow{c}_{\Sigma}(\alpha, k) = \frac{1}{L_{sat}} \left(\frac{\sum_i t_i \, \frac{Q_i u^2_{* i}}{u^2_{* i} - u^2_{th*}}\, c\,(u_{* i}, \alpha_i, k)}{\sum_i t_i}\right) \frac{\overrightarrow{\textbf{k}}}{\|\overrightarrow{\textbf{k}}\|} ,
\label{eq:propagation_sum}
\end{equation}
where $\overrightarrow{\textbf{k}} = (k\, \mathrm{cos}\, \alpha, k\, \mathrm{sin}\, \alpha )$ is the wave vector, $\alpha_i$ is the angle between the perpendicular to each wind and the dune orientation, $t_i$, $Q_i$, and $u_{*i}$ are respectively the time wind blew, the sediment flux and the shear velocity associated with each wind, they act as weighting factors and  $\sigma(u_{* i}, \alpha_i, k)$, and $c\,(u_{* i}, \alpha_i, k)$ are the growth rate and the velocity propagation calculated for each wind (detailed calculation of these variables is described in the Supporting Information). The emerging wavelength and orientation are determined as those which maximise the growth rate.

\section{Results}
\subsection{Wind and sediment transport characteristics}
\begin{figure*}[h!]
\centering
\includegraphics[width=\linewidth]{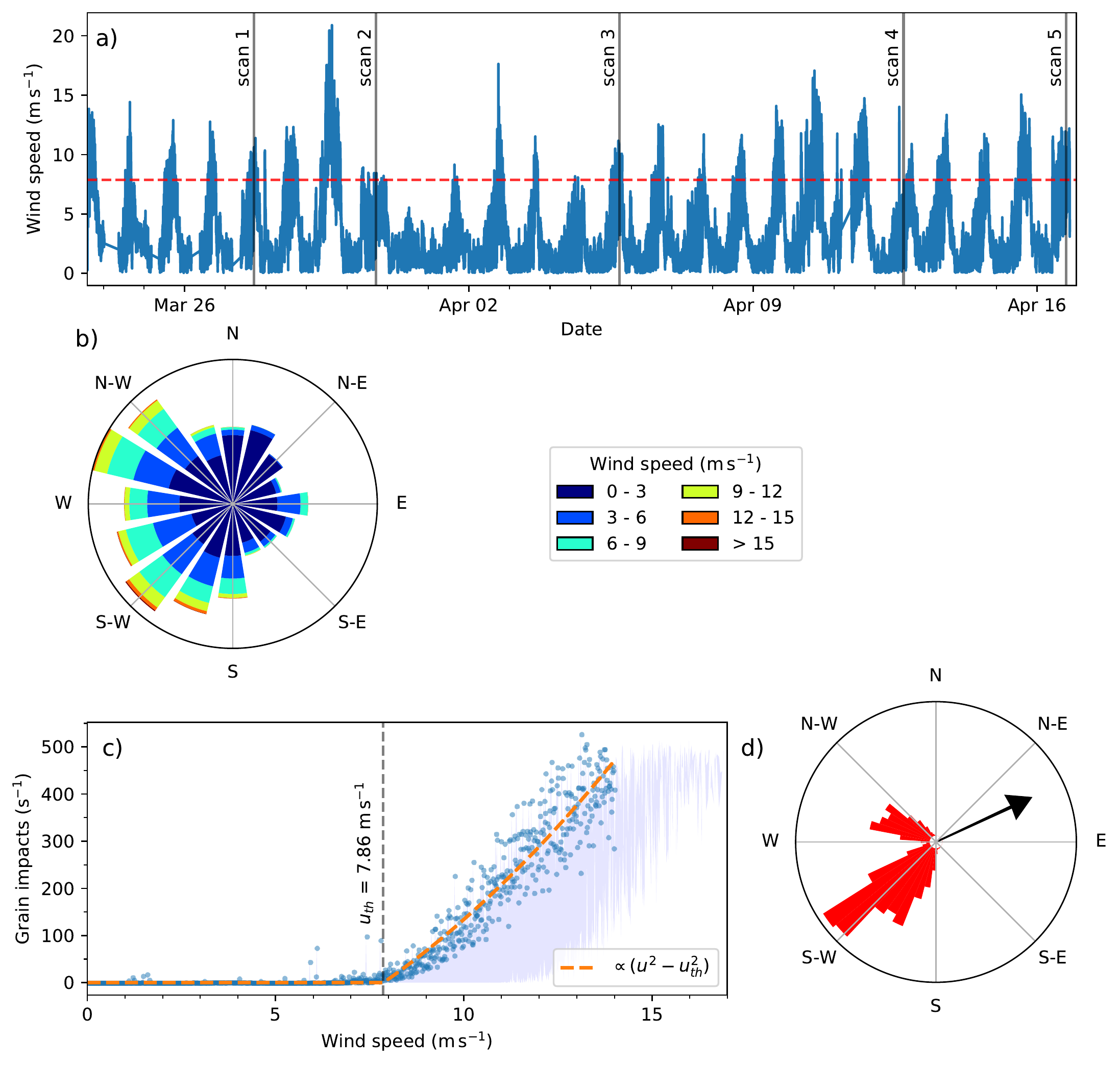}
\caption{a) Raw velocity data measured at z = 3.17 m averaged over 10 seconds. The vertical grey lines represent the date of the TLS scans. The dashed red line is the threshold wind speed $u_{th}$. b) Windrose calculated from the wind speed and direction data measured by the 2D sonic anemometer showing a bimodal wind regime with characteristic winds from 217$\,\pm\,$15$^{\circ}$ and 299$\,\pm\,$10$^{\circ}$. c) Relationship between sand transport (grain impacts recorded by the Sensit) and wind speed. The shaded area represents all data, the dots represent the maximum number of impacts for each 0.01 m$\,$s$^{-1}$ bin of wind speed, the dashed orange line is the fit with a quadratic law. d) Sand flux rose, with sediment flux calculated by the transport law (Equation \ref{eq:transport_law}). The black arrow indicates the Resultant Drift Direction. 
\label{fig:field_data}}
\end{figure*}

During the field campaign the wind speed, measured at a height of 3.70 m, varied between 0 and 20 m$\,$s$^{-1}$ (Fig. \ref{fig:field_data}a), with direction varying mostly from between S and NW.
The stronger, sand moving winds, however, came from two dominant directions, SW and NW (Fig. \ref{fig:field_data}b).

By fitting a quadratic relationship between the wind speed and Sensit output (grain impacts), we determined the threshold wind velocity at 3.17 m height for sediment transport, $u_{\mathrm{th}}$ = 7.86 $\pm$ 1.1 m$\,$s$^{-1}$ (Fig. \ref{fig:field_data}c). Using the law of wall (Equation \ref{eq:law_of_wall}) and assuming an aerodynamic roughness of 0.001 m \cite <approximately 10$\%$ of the height of the sand transport layer,>[]{duran2011aeolian,claudin2013field, gadal2019incipient} we calculated the threshold shear velocity $u_{*\mathrm{th}}$ = 0.39 $\pm$ 0.05 m$\,$s$^{-1}$.
  
Using measured values of sand grain diameter (0.45 $\pm$ 0.07 mm) and density (2650 $\pm$ 54 kg$\,$m$^{-3}$) in Equation \ref{eq:transport_law}, we calculated sediment flux from each wind direction, as shown in Figure \ref{fig:field_data}d. This figure shows that sediment flux originated almost exclusively from two directions, (221$\,\pm\,$15$^{\circ}$ and 303$\,\pm\,$10$^{\circ}$) corresponding with the bimodality evident in the wind directional data. These flux determinations allowed the characterization of the wind regime using DP and RDP (Equations \ref{eq:DP} and \ref{eq:RDP}). For our study, we found $RDP/DP = 0.72$ with a resultant drift direction RDD = 65$^{\circ}$. The $RDP/DP$ ratio indicates a moderate bi-directional wind and sediment transport trend \cite{fryberger1979dune,tsoar2005sand,pearce2005frequency}.

\subsection{Protodune morphodynamics from topographic surveys}

\begin{figure*}[h!]
\centering
\includegraphics[width=.8\linewidth]{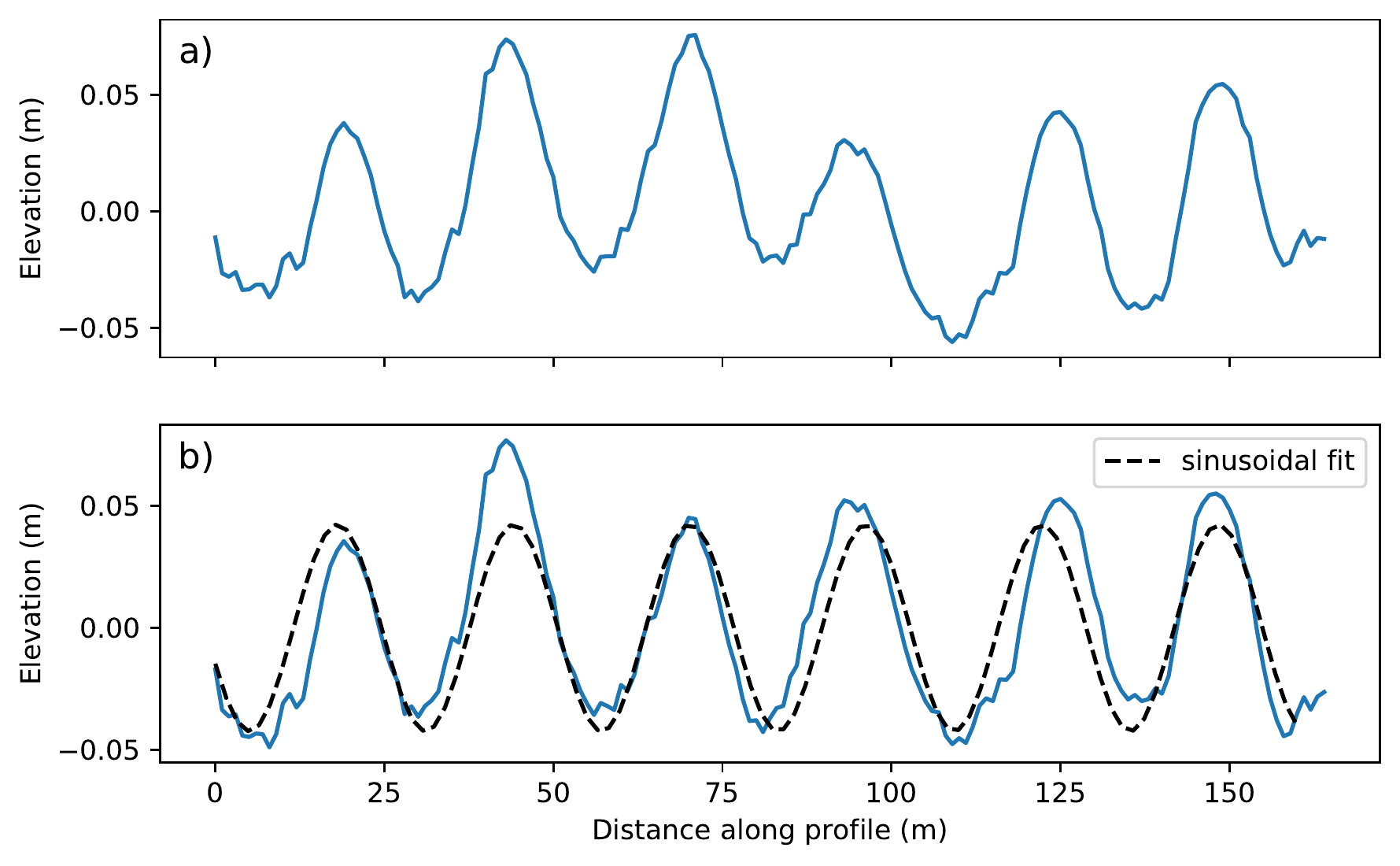}
\caption{a) Example of the raw topographic profile of the 6 studied protodunes (27 March 2019). b) Detrended profiles of the 6 protodunes, the dashed line is a sinusoidal fit. \label{fig:profiles}}
\end{figure*}

From the TLS-derived DEMs we extracted 6 protodune profiles (Fig. \ref{fig:profiles}). The protodunes have very low aspect ratios and do not exhibit an avalanche slope (Fig. \ref{fig:profiles}a). Futhermore, the detrended profile can be approximated by a sin wave (Fig. \ref{fig:profiles}b). These two characteristics are necessary ingredients for the validity of the linear stability analysis \cite{devauchelle2010stability, andreotti2012bedforms,claudin2013field, gadal2019incipient, gadalspatial}.

\begin{figure*}[h!]
\centering
\includegraphics[width=0.9\linewidth]{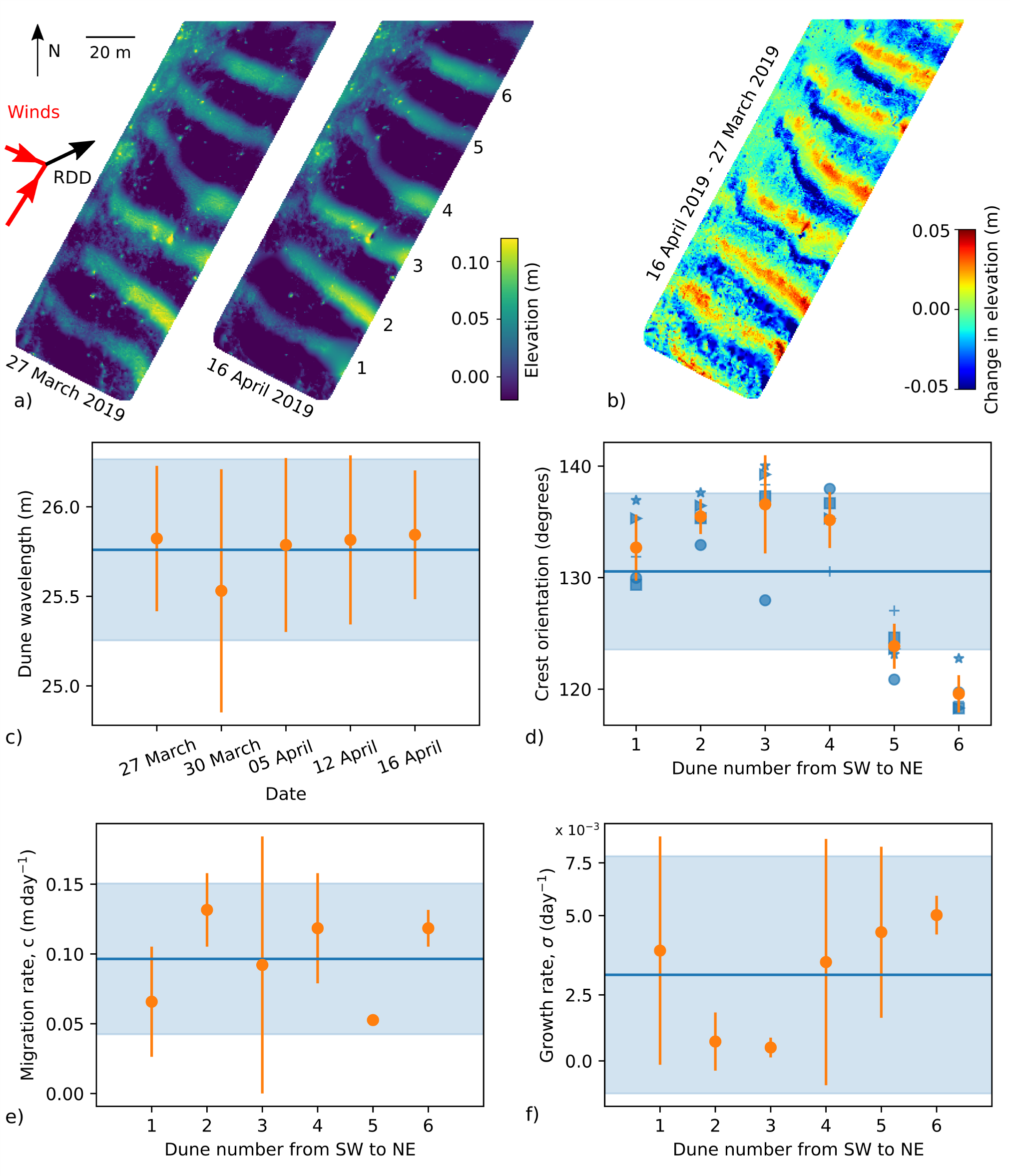}
\caption{a) DEMs of the field area on 27 March 2019 and 16 April 2019, protodunes are labelled from 1 to 6. b) Elevation difference map between 27 March 2019 and 16 April 2019. c) Measured bedform wavelength estimated from each DEM, blue line is the average wavelength, shaded area is the standard deviation. d) Crest orientation of each measured protodune (1 to 6), on each DEM (blue symbol), and average crest orientation (orange dot). e) Measured migration rate of each protodune. f) Measured height increase (i.e. growth rate). Blue lines in c-f represent mean calculated values. \label{fig:DEM}}
\end{figure*}

From the surface elevation data we calculated an average wavelength of 25.7 $\pm$ 0.5~m (Figs. \ref{fig:DEM}a and \ref{fig:DEM}c blue line) and an average orientation of 131 $\pm$ 7$^{\circ}$ for the protodune bedforms (Figs. \ref{fig:DEM}a and \ref{fig:DEM}d blue line). The protodune migration rate over the observation period was 0.096 $\pm$ 0.054 m$\,$day$^{-1}$ (Figs. \ref{fig:DEM}b and \ref{fig:DEM}e blue line). Finally, the increase in protodune amplitude gave a growth rate of, $\sigma$ = 0.003 $\pm$ 0.005 day$^{-1}$ (Figs. \ref{fig:DEM}b and \ref{fig:DEM}f blue line). Within the protodune field there were small variations in wavelength, orientation, growth and migration rates that likely relate to the nuances of the surface sediment availability. No spatial trends in wavelength, orientation, growth and migration rate were observed perpendicular to the channel direction (i.e. between the eastern and western transects) and the overall morphological characteristics of the six bedforms were internally consistent and within the standard deviation of the measurements. 

\subsection{Linear stability analysis}

\begin{figure}[h!]
\centering
\includegraphics[width=.8\linewidth]{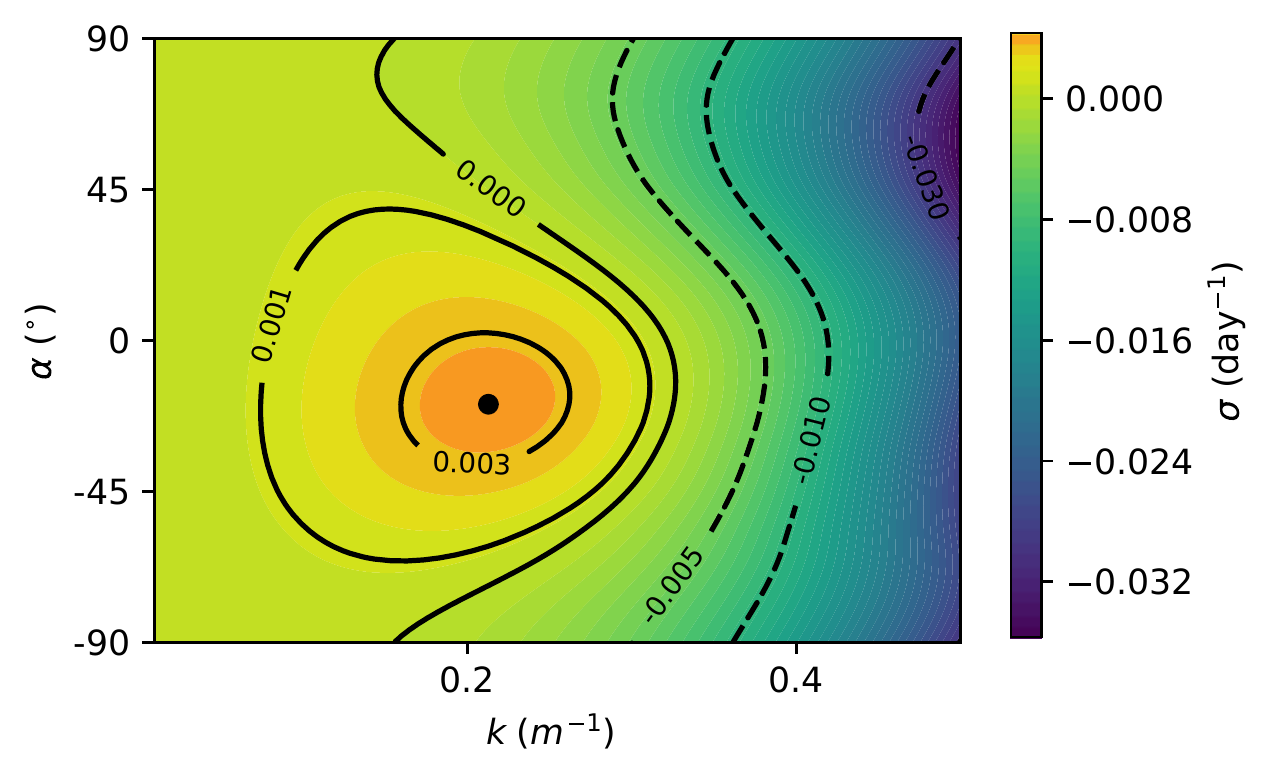}
\caption{\label{fig:growth_rate}  Growth rate $\sigma_{\Sigma}$ as a function of the wavenumber $k$ and the orientation $\alpha$ of the bed perturbation calculated using Equation \ref{eq:growth_sum} for $\cal{A}$ = 3.4 and $\cal{B}$ = 2. The black dot represents the maximum growth rate. }
\end{figure}

Our measurements of wind direction, sediment characteristics and sand flux from the field were used to calculate the most unstable mode of the linear stability analysis. We solved Equation \ref{eq:growth_sum} for values of $k\in[0, 0.5]$ and $\alpha \in [-90^{\circ}, +90^{\circ}]$, with $L_{sat} \approx$ 2 $\pm$ 0.2 m (Equation \ref{eq:L_sat}) and values of $\cal{A}$ and $\cal{B}$ ranging from [3.2, 4] and [1.6, 2.1] respectively (See Supporting Information). We obtained a maximum growth rate of 0.004 $ \pm$ 0.002 day$^{-1}$, reached for a wave number $k_{\mathrm{max}} = 0.23 \pm 0.07$ m$^{-1}$ and an orientation angle $\alpha_{\mathrm{max}} =\, - 19 \pm 4 ^{\circ}$ (Fig. \ref{fig:growth_rate}). This corresponds to a wavelength $\lambda = 2\pi / k = 27 \pm 10\,$m and an orientation of the dune pattern of 136$\pm 4^{\circ}$ (anticlockwise, with 0$^{\circ}$ as North). Additionally, computation of the migration velocity of the most unstable mode using Equation \ref{eq:propagation_sum} and the values of  $k_{\mathrm{max}}$ and  $\alpha_{\mathrm{max}} $ previously determined produces a maximum migration $c$ = 0.12 $\pm$ 0.03~m$\,$day$^{-1}$. 

\section{Discussion}

Our computed statistics resulting from the linear stability analysis of spatial (wavelength and orientation) and temporal (growth rate and migration velocity) characteristics measured in the field can be compared to previous studies that have characterized dune orientation as a function of sand availability and bimodal wind regime \cite{rubin1987bedform,rubin1990flume,werner1999bedform,ping2014emergence,courrech2014two, gao2015phase}. In these studies three parameters are relevant; sand availability, the divergent angle between the two formative winds ($\theta$), and the transport ratio ($N$, the ratio between the sand fluxes that are associated with the two different wind directions) defined as, 
\begin{equation}
N = \frac{Q_1 t_1}{Q_2 t_2}
\end{equation}
where $Q_1$ and $Q_2$ are the sediment fluxes from each direction, and $t_1$ and $t_2$ are the duration of each wind. In the case of infinite sand availability, models such as \citeA{rubin1987bedform} predict that for a divergence angle smaller than 90$^{\circ}$ (as we measured in Sand Creek), the dune crests are more likely to be perpendicular to the resultant sand flux (transverse dunes). In contrast, where the divergent angle is larger than 90$^{\circ}$ \cite <as observed in the Tengger Desert by>[]{ping2014emergence}, the \citeA{rubin1987bedform} model suggests that either longitudinal or oblique dunes form. Where the transport ratio ($N$) is 1, dune crests should align with the resultant sand flux (longitudinal dune), while in the case of a transport ratio higher than 1, the \citeA{rubin1987bedform} model suggests that oblique dunes will form. Subsequent models such as those of \citeA{werner1999bedform} and \citeA{courrech2014two} show a smooth transition between transverse and oblique bedforms, whereas a sharp transition occurs according to the model of \citeA{rubin1987bedform}. The smoothness of this transition depends on the transport ratio, $N$, and the feedback of dune topography on flow strength, computed as the flux-up ratio, $\gamma$, which directly quantifies the increase in sand flux \cite{courrech2014two,gao2015phase}. In the linear regime, dunes do not exhibit an avalanche slope, and the majority of the sediment passing the crest escapes the dune, consequently the flux-up ratio ratio tends toward infinity. The influence of the flux-up ratio is stronger for a divergent angle, $\theta\,\approx$ 90$^{\circ}$.  

In the case of the Sand Creek field site, the mean orientation of the two winds of $\alpha_1 \approx$ 221$ \pm 15^{\circ}$, and $\alpha_2 \approx$ 303 $\pm 10^{\circ}$ creates a divergent angle of $\theta \approx$ 82$^{\circ}$, and we calculated the transport ratio $N$ = 2.1.
Additionally, the protodunes show a smooth profile (without an avalanche slope) so that the equivalent flux-up ratio, $\gamma$ can be considered as infinity \cite{courrech2014two,gao2015phase, gadal2019incipient}. Using the wind values as an input for the model of \citeA{courrech2014two}, we find a dune orientation of 138$^{\circ}$, which corresponds to an angle between the resultant flux and the dune pattern of 73$^{\circ}$. According to the numerical simulation of \citeA{gao2015phase}, an infinite flux-up ratio and a transport ratio higher than 1 smooth the transition from transverse to oblique, consequently the angle between the dune orientation and the resultant sand transport direction moves from 90$^{\circ}$ to an acute angle ($>$ 70$^{\circ}$). Using the linear stability analysis at the Sand Creek field site we calculated an angle between the protodune orientation and the resultant sand transport direction of 71$^{\circ}$, a value within the range of that predicted by the \citeA{gao2015phase} model [70$^{\circ}$, 90$^{\circ}$]. This result complements the work of \citeA{ping2014emergence} as for the first time we observe the formation of oblique dunes under an acute bimodal wind regime.   

To determine the significance and impact of the wind regime bimodality on the pattern of protodune development, we used the linear stability analysis to calculate the morphological and dynamic characteristics of the protodune with the assumption of only a single effective wind direction from the SW. In this uni-directional case the wavelength of the protodunes was calculated to be of the order of 23~m with an orientation of 131$^{\circ}$, celerity 0.22 m$\,$day$^{-1}$, and a growth rate of 0.01 day$^{-1}$. For ease of comparison Table~\ref{table:data} presents a summary of the measured spatial and temporal characteristics of the protodune field and those predicted by the linear stability analysis under both bi-directional and uni-directional wind regimes.

\begin{table}[h]

\caption{\label{table:data}Morphological and dynamical characteristics of the protodune field, measured from the surface DEMs and calculated with the linear stability analysis under a bi-directional and uni-directional wind regime.}
\begin{center}

\begin{tabular}{ccccc}
 
& Wavelengh $\lambda$ & Dune orientation  & Growth rate $\sigma$ & Migration rate $c$ \\
& (m) & (degrees) & (day$^{-1}$) & (m$\,$day$^{-1}$)\\ 
\hline 
\hline 
\begin{tabular}{@{}c@{}c@{}}Morphological\\ measurements from \\ TLS-derived DEM\end{tabular}  & 25.7 $\pm$ 0.5 & 131 $\pm$ 7 & 0.003 $\pm$ 0.005 & 0.096 $\pm$ 0.054 \\ 
\hline 
\begin{tabular}{@{}c@{}}Predictions under \\ a bi-directional wind\end{tabular}  & 27 $\pm$ 10 & 136 $\pm$ 4 & 0.004 $\pm$ 0.002 & 0.12 $\pm$ 0.03 \\ 
\hline 
\begin{tabular}{@{}c@{}}Predictions under a \\ uni-directional SW wind\end{tabular} & 23 $\pm$ 8 & \begin{tabular}{@{}c@{}}perpendicular to the wind  \\ direction (131 $\pm$ 3)\end{tabular}  & 0.01 $\pm$ 0.003 & 0.22 $\pm$ 0.06 \\
\end{tabular} 

\end{center}
\end{table}

This comparison highlights the importance of the secondary wind in the formation of the Sand Creek protodune field.  
As predicted by the analysis of \citeA{gadal2019incipient}, the presence of the NW secondary wind generated a deviation of the flow, consequently a transverse component of the sediment transport appears. The transverse component of the sediment transport has a stabilizing effect which produced an increase in the wavelength, and reduced both the protodune migration and growth rates in comparison with a dominant unimodal wind \cite{gadal2019incipient}.

\section{Conclusion}

Despite recent theoretical and experimental advances in our understanding of protodune initiation, field investigations and tests remain rare and challenging. The experiment presented here is the first field-based study testing the veracity of an extended linear stability analysis for dune initiation in a bimodal wind regime \cite{gadal2019incipient}.

Using measured wind and sand transport data as an input to the linear stability analysis, we found a strong quantitative agreement between the output of the bed instability model and measured field characteristics. Our results validate previous theoretical studies that have highligthted the importance of secondary wind flows on dune orientation \cite{rubin1987bedform,rubin1990flume,werner1999bedform,courrech2014two, gao2015phase}. Further, by simulating the evolution of the same dune field under a uni-directional wind, we show that the presence of a secondary wind in a bimodal regime has the impact of reducing dune growth and migration rates whilst increasing dune wavelength.  

Aside from the quantitative agreement between the theoretical model and the field characteristics provided by this study, our analysis offers confirmation of the validity of the linear stability analysis as a solution for determining patterns of dune emergence and growth on sandy, erodible surfaces. The strong relationship between wind regime and dune morphology substantiated in this study also offers confidence in analyses of the reverse problem, where determination of the wind regime in a natural environment is required based on evidence provided by dune morphology. Such a requirement is widespread in analyses of extra-terrestrial dune surfaces where wind data are absent \cite <e.g.>[]{jia2017giant,fernandez2018first, vinent2019unified}.

Whilst our analysis provides the first field validation of the linear stability analysis in bimodal wind regimes, an important step in futhering our understanding of dune initiation processes, it suffers from some restrictions due to the lack of field testing of key model parameters. In our study, the parameters $\cal{A}$ and $\cal{B}$, required adjustment to obtain agreement between the outputs of the linear stability model and the measured morphological characteristics. Our understanding of these parameters needs to be quantitatively strengthened with further field and experimental data focusing on measurements of the spatial shift between the maximum shear velocity and the crest. Currently, the only data available for the determination of $\cal{A}$ and $\cal{B}$ are from the field measurements of \citeA{claudin2013field}. Further, the calculation of the saturation length $L_{\mathrm{sat}}$ requires additional field quantification with measurements of the spatial shift between maximum shear velocity and resulting maximum sand flux. With the addition of such field and experimental data to refinements of the linear stability analysis, we are confident that our understanding of dune initiation processes in complex and natural wind regimes can be significantly advanced.

%

\acknowledgments
This work was funded by the TOAD (The Origin of Aeolian Dunes) project (Natural Environment Research Council, UK and National Science Foundation, USA; NE/R010196NSFGEO-NERC, NSF-GEO-1829541 and NSF-GEO-1829513). Research was undertaken at GSD under a Scientific Research and Collection permit GRSA-2018-SCI-004 and we are very grateful for field, access and logistical support from F. Bunch and other park staff. Data processing used IRIDIS Southampton Computing Facility. We thank R. Ewing for the use of his Sensit instrument, G. Kocurek, R. Reynolds, A. Gunn and R. Ewing for field assistance and C. Gadal and C. Narteau for fruitful discussions about the linear stability analysis. J.M. Nield gratefully acknowledges the Department of Geology and Geophysics, Texas A$\&$M University, where she was supported by a Michel T Halbouty Visiting Chair during the field campaign. The data used in this manuscript can be found in the NERC National Geological Data Center ( https://doi.org/10.5285/46af71db-1caa-4c88-bd6a-c9724dd90df8). Supplementary figures and text can be found in the supporting information. We thank the anonymous reviewers for their insightful comments and suggestions.


%
%

\bibliography{bimodal}

\begin{thebibliography}{}

\bibitem [\protect \citeauthoryear {%
Andreotti%
}{%
Andreotti%
}{%
{\protect \APACyear {2004}}%
}]{%
andreotti2004two}
\APACinsertmetastar {%
andreotti2004two}%
\begin{APACrefauthors}%
Andreotti, B.%
\end{APACrefauthors}%
\unskip\
\newblock
\APACrefYearMonthDay{2004}{}{}.
\newblock
{\BBOQ}\APACrefatitle {A two-species model of aeolian sand transport} {A
  two-species model of aeolian sand transport}.{\BBCQ}
\newblock
\APACjournalVolNumPages{Journal of Fluid Mechanics}{510}{}{47--70}.
\PrintBackRefs{\CurrentBib}

\bibitem [\protect \citeauthoryear {%
Andreotti%
, Claudin%
, Devauchelle%
, Dur{\'a}n%
\BCBL {}\ \BBA {} Fourri{\`e}re%
}{%
Andreotti%
\ \protect \BOthers {.}}{%
{\protect \APACyear {2012}}%
}]{%
andreotti2012bedforms}
\APACinsertmetastar {%
andreotti2012bedforms}%
\begin{APACrefauthors}%
Andreotti, B.%
, Claudin, P.%
, Devauchelle, O.%
, Dur{\'a}n, O.%
\BCBL {}\ \BBA {} Fourri{\`e}re, A.%
\end{APACrefauthors}%
\unskip\
\newblock
\APACrefYearMonthDay{2012}{}{}.
\newblock
{\BBOQ}\APACrefatitle {Bedforms in a turbulent stream: ripples, chevrons and
  antidunes} {Bedforms in a turbulent stream: ripples, chevrons and
  antidunes}.{\BBCQ}
\newblock
\APACjournalVolNumPages{Journal of Fluid Mechanics}{690}{}{94--128}.
\PrintBackRefs{\CurrentBib}

\bibitem [\protect \citeauthoryear {%
Andreotti%
, Claudin%
\BCBL {}\ \BBA {} Douady%
}{%
Andreotti%
\ \protect \BOthers {.}}{%
{\protect \APACyear {2002}}%
}]{%
andreotti2002selection}
\APACinsertmetastar {%
andreotti2002selection}%
\begin{APACrefauthors}%
Andreotti, B.%
, Claudin, P.%
\BCBL {}\ \BBA {} Douady, S.%
\end{APACrefauthors}%
\unskip\
\newblock
\APACrefYearMonthDay{2002}{}{}.
\newblock
{\BBOQ}\APACrefatitle {Selection of dune shapes and velocities Part 2: A
  two-dimensional modelling} {Selection of dune shapes and velocities part 2: A
  two-dimensional modelling}.{\BBCQ}
\newblock
\APACjournalVolNumPages{The European Physical Journal B-Condensed Matter and
  Complex Systems}{28}{3}{341--352}.
\PrintBackRefs{\CurrentBib}

\bibitem [\protect \citeauthoryear {%
Andreotti%
, Claudin%
\BCBL {}\ \BBA {} Pouliquen%
}{%
Andreotti%
\ \protect \BOthers {.}}{%
{\protect \APACyear {2010}}%
}]{%
andreotti2010measurements}
\APACinsertmetastar {%
andreotti2010measurements}%
\begin{APACrefauthors}%
Andreotti, B.%
, Claudin, P.%
\BCBL {}\ \BBA {} Pouliquen, O.%
\end{APACrefauthors}%
\unskip\
\newblock
\APACrefYearMonthDay{2010}{}{}.
\newblock
{\BBOQ}\APACrefatitle {Measurements of the aeolian sand transport saturation
  length} {Measurements of the aeolian sand transport saturation
  length}.{\BBCQ}
\newblock
\APACjournalVolNumPages{Geomorphology}{123}{3-4}{343--348}.
\PrintBackRefs{\CurrentBib}

\bibitem [\protect \citeauthoryear {%
Baddock%
, Nield%
\BCBL {}\ \BBA {} Wiggs%
}{%
Baddock%
\ \protect \BOthers {.}}{%
{\protect \APACyear {2018}}%
}]{%
baddock2018early}
\APACinsertmetastar {%
baddock2018early}%
\begin{APACrefauthors}%
Baddock, M\BPBI C.%
, Nield, J\BPBI M.%
\BCBL {}\ \BBA {} Wiggs, G\BPBI F\BPBI S.%
\end{APACrefauthors}%
\unskip\
\newblock
\APACrefYearMonthDay{2018}{}{}.
\newblock
{\BBOQ}\APACrefatitle {Early-stage aeolian protodunes: Bedform development and
  sand transport dynamics} {Early-stage aeolian protodunes: Bedform development
  and sand transport dynamics}.{\BBCQ}
\newblock
\APACjournalVolNumPages{Earth Surface Processes and
  Landforms}{43}{1}{339--346}.
\PrintBackRefs{\CurrentBib}

\bibitem [\protect \citeauthoryear {%
Bagnold%
}{%
Bagnold%
}{%
{\protect \APACyear {1937}}%
}]{%
bagnold1937transport}
\APACinsertmetastar {%
bagnold1937transport}%
\begin{APACrefauthors}%
Bagnold, R\BPBI A.%
\end{APACrefauthors}%
\unskip\
\newblock
\APACrefYearMonthDay{1937}{}{}.
\newblock
{\BBOQ}\APACrefatitle {The transport of sand by wind} {The transport of sand by
  wind}.{\BBCQ}
\newblock
\APACjournalVolNumPages{The Geographical Journal}{89}{5}{409--438}.
\PrintBackRefs{\CurrentBib}

\bibitem [\protect \citeauthoryear {%
Bullard%
, Thomas%
, Livingstone%
\BCBL {}\ \BBA {} Wiggs%
}{%
Bullard%
\ \protect \BOthers {.}}{%
{\protect \APACyear {1996}}%
}]{%
bullard1996wind}
\APACinsertmetastar {%
bullard1996wind}%
\begin{APACrefauthors}%
Bullard, J.%
, Thomas, D.%
, Livingstone, I.%
\BCBL {}\ \BBA {} Wiggs, G\BPBI F\BPBI S.%
\end{APACrefauthors}%
\unskip\
\newblock
\APACrefYearMonthDay{1996}{}{}.
\newblock
{\BBOQ}\APACrefatitle {Wind energy variations in the southwestern Kalahari
  Desert and implications for linear dunefield activity} {Wind energy
  variations in the southwestern kalahari desert and implications for linear
  dunefield activity}.{\BBCQ}
\newblock
\APACjournalVolNumPages{Earth Surface Processes and
  Landforms}{21}{3}{263--278}.
\PrintBackRefs{\CurrentBib}

\bibitem [\protect \citeauthoryear {%
Charru%
, Andreotti%
\BCBL {}\ \BBA {} Claudin%
}{%
Charru%
\ \protect \BOthers {.}}{%
{\protect \APACyear {2013}}%
}]{%
charru2013sand}
\APACinsertmetastar {%
charru2013sand}%
\begin{APACrefauthors}%
Charru, F.%
, Andreotti, B.%
\BCBL {}\ \BBA {} Claudin, P.%
\end{APACrefauthors}%
\unskip\
\newblock
\APACrefYearMonthDay{2013}{}{}.
\newblock
{\BBOQ}\APACrefatitle {Sand ripples and dunes} {Sand ripples and dunes}.{\BBCQ}
\newblock
\APACjournalVolNumPages{Annual Review of Fluid Mechanics}{45}{}{469--493}.
\PrintBackRefs{\CurrentBib}

\bibitem [\protect \citeauthoryear {%
Claudin%
\ \BBA {} Andreotti%
}{%
Claudin%
\ \BBA {} Andreotti%
}{%
{\protect \APACyear {2006}}%
}]{%
claudin2006scaling}
\APACinsertmetastar {%
claudin2006scaling}%
\begin{APACrefauthors}%
Claudin, P.%
\BCBT {}\ \BBA {} Andreotti, B.%
\end{APACrefauthors}%
\unskip\
\newblock
\APACrefYearMonthDay{2006}{}{}.
\newblock
{\BBOQ}\APACrefatitle {A scaling law for aeolian dunes on Mars, Venus, Earth,
  and for subaqueous ripples} {A scaling law for aeolian dunes on mars, venus,
  earth, and for subaqueous ripples}.{\BBCQ}
\newblock
\APACjournalVolNumPages{Earth and Planetary Science Letters}{252}{1-2}{30--44}.
\PrintBackRefs{\CurrentBib}

\bibitem [\protect \citeauthoryear {%
Claudin%
, Wiggs%
\BCBL {}\ \BBA {} Andreotti%
}{%
Claudin%
\ \protect \BOthers {.}}{%
{\protect \APACyear {2013}}%
}]{%
claudin2013field}
\APACinsertmetastar {%
claudin2013field}%
\begin{APACrefauthors}%
Claudin, P.%
, Wiggs, G\BPBI F\BPBI S.%
\BCBL {}\ \BBA {} Andreotti, B.%
\end{APACrefauthors}%
\unskip\
\newblock
\APACrefYearMonthDay{2013}{}{}.
\newblock
{\BBOQ}\APACrefatitle {Field evidence for the upwind velocity shift at the
  crest of low dunes} {Field evidence for the upwind velocity shift at the
  crest of low dunes}.{\BBCQ}
\newblock
\APACjournalVolNumPages{Boundary-layer meteorology}{148}{1}{195--206}.
\PrintBackRefs{\CurrentBib}

\bibitem [\protect \citeauthoryear {%
Colombini%
}{%
Colombini%
}{%
{\protect \APACyear {2004}}%
}]{%
colombini2004revisiting}
\APACinsertmetastar {%
colombini2004revisiting}%
\begin{APACrefauthors}%
Colombini, M.%
\end{APACrefauthors}%
\unskip\
\newblock
\APACrefYearMonthDay{2004}{}{}.
\newblock
{\BBOQ}\APACrefatitle {Revisiting the linear theory of sand dune formation}
  {Revisiting the linear theory of sand dune formation}.{\BBCQ}
\newblock
\APACjournalVolNumPages{Journal of Fluid Mechanics}{502}{}{1--16}.
\PrintBackRefs{\CurrentBib}

\bibitem [\protect \citeauthoryear {%
Courrech~du Pont%
, Narteau%
\BCBL {}\ \BBA {} Gao%
}{%
Courrech~du Pont%
\ \protect \BOthers {.}}{%
{\protect \APACyear {2014}}%
}]{%
courrech2014two}
\APACinsertmetastar {%
courrech2014two}%
\begin{APACrefauthors}%
Courrech~du Pont, S.%
, Narteau, C.%
\BCBL {}\ \BBA {} Gao, X.%
\end{APACrefauthors}%
\unskip\
\newblock
\APACrefYearMonthDay{2014}{}{}.
\newblock
{\BBOQ}\APACrefatitle {Two modes for dune orientation} {Two modes for dune
  orientation}.{\BBCQ}
\newblock
\APACjournalVolNumPages{Geology}{42}{9}{743--746}.
\PrintBackRefs{\CurrentBib}

\bibitem [\protect \citeauthoryear {%
Creyssels%
\ \protect \BOthers {.}}{%
Creyssels%
\ \protect \BOthers {.}}{%
{\protect \APACyear {2009}}%
}]{%
creyssels2009saltating}
\APACinsertmetastar {%
creyssels2009saltating}%
\begin{APACrefauthors}%
Creyssels, M.%
, Dupont, P.%
, El~Moctar, A\BPBI O.%
, Valance, A.%
, Cantat, I.%
, Jenkins, J\BPBI T.%
\BDBL {}Rasmussen, K\BPBI R.%
\end{APACrefauthors}%
\unskip\
\newblock
\APACrefYearMonthDay{2009}{}{}.
\newblock
{\BBOQ}\APACrefatitle {Saltating particles in a turbulent boundary layer:
  experiment and theory} {Saltating particles in a turbulent boundary layer:
  experiment and theory}.{\BBCQ}
\newblock
\APACjournalVolNumPages{Journal of Fluid Mechanics}{625}{}{47}.
\PrintBackRefs{\CurrentBib}

\bibitem [\protect \citeauthoryear {%
Devauchelle%
\ \protect \BOthers {.}}{%
Devauchelle%
\ \protect \BOthers {.}}{%
{\protect \APACyear {2010}}%
}]{%
devauchelle2010stability}
\APACinsertmetastar {%
devauchelle2010stability}%
\begin{APACrefauthors}%
Devauchelle, O.%
, Malverti, L.%
, Lajeunesse, E.%
, Lagr{\'e}e, P\BHBI Y.%
, Josserand, C.%
\BCBL {}\ \BBA {} Thu-Lam, K\BHBI D\BPBI N.%
\end{APACrefauthors}%
\unskip\
\newblock
\APACrefYearMonthDay{2010}{}{}.
\newblock
{\BBOQ}\APACrefatitle {Stability of bedforms in laminar flows with free
  surface: from bars to ripples} {Stability of bedforms in laminar flows with
  free surface: from bars to ripples}.{\BBCQ}
\newblock
\APACjournalVolNumPages{Journal of Fluid Mechanics}{642}{}{329--348}.
\PrintBackRefs{\CurrentBib}

\bibitem [\protect \citeauthoryear {%
Dur{\'a}n%
, Claudin%
\BCBL {}\ \BBA {} Andreotti%
}{%
Dur{\'a}n%
\ \protect \BOthers {.}}{%
{\protect \APACyear {2011}}%
}]{%
duran2011aeolian}
\APACinsertmetastar {%
duran2011aeolian}%
\begin{APACrefauthors}%
Dur{\'a}n, O.%
, Claudin, P.%
\BCBL {}\ \BBA {} Andreotti, B.%
\end{APACrefauthors}%
\unskip\
\newblock
\APACrefYearMonthDay{2011}{}{}.
\newblock
{\BBOQ}\APACrefatitle {On aeolian transport: Grain-scale interactions,
  dynamical mechanisms and scaling laws} {On aeolian transport: Grain-scale
  interactions, dynamical mechanisms and scaling laws}.{\BBCQ}
\newblock
\APACjournalVolNumPages{Aeolian Research}{3}{3}{243--270}.
\PrintBackRefs{\CurrentBib}

\bibitem [\protect \citeauthoryear {%
Dur{\'a}n~Vinent%
, Andreotti%
, Claudin%
\BCBL {}\ \BBA {} Winter%
}{%
Dur{\'a}n~Vinent%
\ \protect \BOthers {.}}{%
{\protect \APACyear {2019}}%
}]{%
vinent2019unified}
\APACinsertmetastar {%
vinent2019unified}%
\begin{APACrefauthors}%
Dur{\'a}n~Vinent, O.%
, Andreotti, B.%
, Claudin, P.%
\BCBL {}\ \BBA {} Winter, C.%
\end{APACrefauthors}%
\unskip\
\newblock
\APACrefYearMonthDay{2019}{}{}.
\newblock
{\BBOQ}\APACrefatitle {A unified model of ripples and dunes in water and
  planetary environments} {A unified model of ripples and dunes in water and
  planetary environments}.{\BBCQ}
\newblock
\APACjournalVolNumPages{Nature Geoscience}{12}{5}{345--350}.
\PrintBackRefs{\CurrentBib}

\bibitem [\protect \citeauthoryear {%
Eastwood%
, Nield%
, Baas%
\BCBL {}\ \BBA {} Kocurek%
}{%
Eastwood%
\ \protect \BOthers {.}}{%
{\protect \APACyear {2011}}%
}]{%
eastwood2011modelling}
\APACinsertmetastar {%
eastwood2011modelling}%
\begin{APACrefauthors}%
Eastwood, E.%
, Nield, J.%
, Baas, A.%
\BCBL {}\ \BBA {} Kocurek, G.%
\end{APACrefauthors}%
\unskip\
\newblock
\APACrefYearMonthDay{2011}{}{}.
\newblock
{\BBOQ}\APACrefatitle {Modelling controls on aeolian dune-field pattern
  evolution} {Modelling controls on aeolian dune-field pattern
  evolution}.{\BBCQ}
\newblock
\APACjournalVolNumPages{Sedimentology}{58}{6}{1391--1406}.
\PrintBackRefs{\CurrentBib}

\bibitem [\protect \citeauthoryear {%
Elbelrhiti%
, Claudin%
\BCBL {}\ \BBA {} Andreotti%
}{%
Elbelrhiti%
\ \protect \BOthers {.}}{%
{\protect \APACyear {2005}}%
}]{%
elbelrhiti2005field}
\APACinsertmetastar {%
elbelrhiti2005field}%
\begin{APACrefauthors}%
Elbelrhiti, H.%
, Claudin, P.%
\BCBL {}\ \BBA {} Andreotti, B.%
\end{APACrefauthors}%
\unskip\
\newblock
\APACrefYearMonthDay{2005}{}{}.
\newblock
{\BBOQ}\APACrefatitle {Field evidence for surface-wave-induced instability of
  sand dunes} {Field evidence for surface-wave-induced instability of sand
  dunes}.{\BBCQ}
\newblock
\APACjournalVolNumPages{Nature}{437}{7059}{720}.
\PrintBackRefs{\CurrentBib}

\bibitem [\protect \citeauthoryear {%
Fernandez-Cascales%
\ \protect \BOthers {.}}{%
Fernandez-Cascales%
\ \protect \BOthers {.}}{%
{\protect \APACyear {2018}}%
}]{%
fernandez2018first}
\APACinsertmetastar {%
fernandez2018first}%
\begin{APACrefauthors}%
Fernandez-Cascales, L.%
, Lucas, A.%
, Rodriguez, S.%
, Gao, X.%
, Spiga, A.%
\BCBL {}\ \BBA {} Narteau, C.%
\end{APACrefauthors}%
\unskip\
\newblock
\APACrefYearMonthDay{2018}{}{}.
\newblock
{\BBOQ}\APACrefatitle {First quantification of relationship between dune
  orientation and sediment availability, Olympia Undae, Mars} {First
  quantification of relationship between dune orientation and sediment
  availability, olympia undae, mars}.{\BBCQ}
\newblock
\APACjournalVolNumPages{Earth and Planetary Science Letters}{489}{}{241--250}.
\PrintBackRefs{\CurrentBib}

\bibitem [\protect \citeauthoryear {%
Fourri{\`e}re%
, Claudin%
\BCBL {}\ \BBA {} Andreotti%
}{%
Fourri{\`e}re%
\ \protect \BOthers {.}}{%
{\protect \APACyear {2010}}%
}]{%
fourriere2010bedforms}
\APACinsertmetastar {%
fourriere2010bedforms}%
\begin{APACrefauthors}%
Fourri{\`e}re, A.%
, Claudin, P.%
\BCBL {}\ \BBA {} Andreotti, B.%
\end{APACrefauthors}%
\unskip\
\newblock
\APACrefYearMonthDay{2010}{}{}.
\newblock
{\BBOQ}\APACrefatitle {Bedforms in a turbulent stream: formation of ripples by
  primary linear instability and of dunes by nonlinear pattern coarsening}
  {Bedforms in a turbulent stream: formation of ripples by primary linear
  instability and of dunes by nonlinear pattern coarsening}.{\BBCQ}
\newblock
\APACjournalVolNumPages{Journal of Fluid Mechanics}{649}{}{287--328}.
\PrintBackRefs{\CurrentBib}

\bibitem [\protect \citeauthoryear {%
Fryberger%
\ \BBA {} Dean%
}{%
Fryberger%
\ \BBA {} Dean%
}{%
{\protect \APACyear {1979}}%
}]{%
fryberger1979dune}
\APACinsertmetastar {%
fryberger1979dune}%
\begin{APACrefauthors}%
Fryberger, S\BPBI G.%
\BCBT {}\ \BBA {} Dean, G.%
\end{APACrefauthors}%
\unskip\
\newblock
\APACrefYearMonthDay{1979}{}{}.
\newblock
{\BBOQ}\APACrefatitle {Dune forms and wind regime} {Dune forms and wind
  regime}.{\BBCQ}
\newblock
\BIn{} \APACrefbtitle {A study of global sand seas} {A study of global sand
  seas}\ (\BVOL\ 1052, \BPGS\ 137--169).
\newblock
\APACaddressPublisher{}{US Government Printing Office Washington}.
\PrintBackRefs{\CurrentBib}

\bibitem [\protect \citeauthoryear {%
Gadal%
, Narteau%
, du Pont%
, Rozier%
\BCBL {}\ \BBA {} Claudin%
}{%
Gadal%
\ \protect \BOthers {.}}{%
{\protect \APACyear {2019}}%
}]{%
gadal2019incipient}
\APACinsertmetastar {%
gadal2019incipient}%
\begin{APACrefauthors}%
Gadal, C.%
, Narteau, C.%
, du Pont, S\BPBI C.%
, Rozier, O.%
\BCBL {}\ \BBA {} Claudin, P.%
\end{APACrefauthors}%
\unskip\
\newblock
\APACrefYearMonthDay{2019}{}{}.
\newblock
{\BBOQ}\APACrefatitle {Incipient bedforms in a bidirectional wind regime}
  {Incipient bedforms in a bidirectional wind regime}.{\BBCQ}
\newblock
\APACjournalVolNumPages{Journal of Fluid Mechanics}{862}{}{490--516}.
\PrintBackRefs{\CurrentBib}

\bibitem [\protect \citeauthoryear {%
Gadal%
\ \protect \BOthers {.}}{%
Gadal%
\ \protect \BOthers {.}}{%
{\protect \APACyear {2020}}%
}]{%
gadalspatial}
\APACinsertmetastar {%
gadalspatial}%
\begin{APACrefauthors}%
Gadal, C.%
, Narteau, C.%
, Ewing, R.%
, Gunn, A.%
, Jerolmack, D.%
, Andreotti, B.%
\BCBL {}\ \BBA {} Claudin, P.%
\end{APACrefauthors}%
\unskip\
\newblock
\APACrefYearMonthDay{2020}{}{}.
\newblock
{\BBOQ}\APACrefatitle {Spatial and temporal development of incipient dunes}
  {Spatial and temporal development of incipient dunes}.{\BBCQ}
\newblock
\APACjournalVolNumPages{Geophysical Research Letters}{}{}{e2020GL088919}.
\PrintBackRefs{\CurrentBib}

\bibitem [\protect \citeauthoryear {%
Gao%
, Gadal%
, Rozier%
\BCBL {}\ \BBA {} Narteau%
}{%
Gao%
\ \protect \BOthers {.}}{%
{\protect \APACyear {2018}}%
}]{%
gao2018morphodynamics}
\APACinsertmetastar {%
gao2018morphodynamics}%
\begin{APACrefauthors}%
Gao, X.%
, Gadal, C.%
, Rozier, O.%
\BCBL {}\ \BBA {} Narteau, C.%
\end{APACrefauthors}%
\unskip\
\newblock
\APACrefYearMonthDay{2018}{}{}.
\newblock
{\BBOQ}\APACrefatitle {Morphodynamics of barchan and dome dunes under variable
  wind regimes} {Morphodynamics of barchan and dome dunes under variable wind
  regimes}.{\BBCQ}
\newblock
\APACjournalVolNumPages{Geology}{46}{9}{743--746}.
\PrintBackRefs{\CurrentBib}

\bibitem [\protect \citeauthoryear {%
Gao%
, Narteau%
, Rozier%
\BCBL {}\ \BBA {} Courrech~du Pont%
}{%
Gao%
\ \protect \BOthers {.}}{%
{\protect \APACyear {2015}}%
}]{%
gao2015phase}
\APACinsertmetastar {%
gao2015phase}%
\begin{APACrefauthors}%
Gao, X.%
, Narteau, C.%
, Rozier, O.%
\BCBL {}\ \BBA {} Courrech~du Pont, S.%
\end{APACrefauthors}%
\unskip\
\newblock
\APACrefYearMonthDay{2015}{}{}.
\newblock
{\BBOQ}\APACrefatitle {Phase diagrams of dune shape and orientation depending
  on sand availability} {Phase diagrams of dune shape and orientation depending
  on sand availability}.{\BBCQ}
\newblock
\APACjournalVolNumPages{Scientific reports}{5}{1}{1--12}.
\PrintBackRefs{\CurrentBib}

\bibitem [\protect \citeauthoryear {%
Gillette%
\ \BBA {} Stockton%
}{%
Gillette%
\ \BBA {} Stockton%
}{%
{\protect \APACyear {1986}}%
}]{%
gillette1986mass}
\APACinsertmetastar {%
gillette1986mass}%
\begin{APACrefauthors}%
Gillette, D\BPBI A.%
\BCBT {}\ \BBA {} Stockton, P.%
\end{APACrefauthors}%
\unskip\
\newblock
\APACrefYearMonthDay{1986}{}{}.
\newblock
{\BBOQ}\APACrefatitle {Mass momentum and kinetic energy fluxes of saltating
  particles} {Mass momentum and kinetic energy fluxes of saltating
  particles}.{\BBCQ}
\newblock
\BIn{} \APACrefbtitle {Annual symposium of geomorphology. 17} {Annual symposium
  of geomorphology. 17}\ (\BPGS\ 35--56).
\PrintBackRefs{\CurrentBib}

\bibitem [\protect \citeauthoryear {%
Hersen%
, Douady%
\BCBL {}\ \BBA {} Andreotti%
}{%
Hersen%
\ \protect \BOthers {.}}{%
{\protect \APACyear {2002}}%
}]{%
hersen2002relevant}
\APACinsertmetastar {%
hersen2002relevant}%
\begin{APACrefauthors}%
Hersen, P.%
, Douady, S.%
\BCBL {}\ \BBA {} Andreotti, B.%
\end{APACrefauthors}%
\unskip\
\newblock
\APACrefYearMonthDay{2002}{}{}.
\newblock
{\BBOQ}\APACrefatitle {Relevant length scale of barchan dunes} {Relevant length
  scale of barchan dunes}.{\BBCQ}
\newblock
\APACjournalVolNumPages{Physical Review Letters}{89}{26}{264301}.
\PrintBackRefs{\CurrentBib}

\bibitem [\protect \citeauthoryear {%
Ho%
, Valance%
, Dupont%
\BCBL {}\ \BBA {} El~Moctar%
}{%
Ho%
\ \protect \BOthers {.}}{%
{\protect \APACyear {2011}}%
}]{%
ho2011scaling}
\APACinsertmetastar {%
ho2011scaling}%
\begin{APACrefauthors}%
Ho, T\BPBI D.%
, Valance, A.%
, Dupont, P.%
\BCBL {}\ \BBA {} El~Moctar, A\BPBI O.%
\end{APACrefauthors}%
\unskip\
\newblock
\APACrefYearMonthDay{2011}{}{}.
\newblock
{\BBOQ}\APACrefatitle {Scaling laws in aeolian sand transport} {Scaling laws in
  aeolian sand transport}.{\BBCQ}
\newblock
\APACjournalVolNumPages{Physical Review Letters}{106}{9}{094501}.
\PrintBackRefs{\CurrentBib}

\bibitem [\protect \citeauthoryear {%
Hugenholtz%
\ \BBA {} Barchyn%
}{%
Hugenholtz%
\ \BBA {} Barchyn%
}{%
{\protect \APACyear {2010}}%
}]{%
hugenholtz2010spatial}
\APACinsertmetastar {%
hugenholtz2010spatial}%
\begin{APACrefauthors}%
Hugenholtz, C\BPBI H.%
\BCBT {}\ \BBA {} Barchyn, T\BPBI E.%
\end{APACrefauthors}%
\unskip\
\newblock
\APACrefYearMonthDay{2010}{}{}.
\newblock
{\BBOQ}\APACrefatitle {Spatial analysis of sand dunes with a new global
  topographic dataset: new approaches and opportunities} {Spatial analysis of
  sand dunes with a new global topographic dataset: new approaches and
  opportunities}.{\BBCQ}
\newblock
\APACjournalVolNumPages{Earth surface processes and
  landforms}{35}{8}{986--992}.
\PrintBackRefs{\CurrentBib}

\bibitem [\protect \citeauthoryear {%
Iversen%
\ \BBA {} Rasmussen%
}{%
Iversen%
\ \BBA {} Rasmussen%
}{%
{\protect \APACyear {1999}}%
}]{%
iversen1999effect}
\APACinsertmetastar {%
iversen1999effect}%
\begin{APACrefauthors}%
Iversen, J\BPBI D.%
\BCBT {}\ \BBA {} Rasmussen, K\BPBI R.%
\end{APACrefauthors}%
\unskip\
\newblock
\APACrefYearMonthDay{1999}{}{}.
\newblock
{\BBOQ}\APACrefatitle {The effect of wind speed and bed slope on sand
  transport} {The effect of wind speed and bed slope on sand transport}.{\BBCQ}
\newblock
\APACjournalVolNumPages{Sedimentology}{46}{4}{723--731}.
\PrintBackRefs{\CurrentBib}

\bibitem [\protect \citeauthoryear {%
Jackson%
\ \BBA {} Hunt%
}{%
Jackson%
\ \BBA {} Hunt%
}{%
{\protect \APACyear {1975}}%
}]{%
jackson1975turbulent}
\APACinsertmetastar {%
jackson1975turbulent}%
\begin{APACrefauthors}%
Jackson, P.%
\BCBT {}\ \BBA {} Hunt, J.%
\end{APACrefauthors}%
\unskip\
\newblock
\APACrefYearMonthDay{1975}{}{}.
\newblock
{\BBOQ}\APACrefatitle {Turbulent wind flow over a low hill} {Turbulent wind
  flow over a low hill}.{\BBCQ}
\newblock
\APACjournalVolNumPages{Quarterly Journal of the Royal Meteorological
  Society}{101}{430}{929--955}.
\PrintBackRefs{\CurrentBib}

\bibitem [\protect \citeauthoryear {%
Jia%
, Andreotti%
\BCBL {}\ \BBA {} Claudin%
}{%
Jia%
\ \protect \BOthers {.}}{%
{\protect \APACyear {2017}}%
}]{%
jia2017giant}
\APACinsertmetastar {%
jia2017giant}%
\begin{APACrefauthors}%
Jia, P.%
, Andreotti, B.%
\BCBL {}\ \BBA {} Claudin, P.%
\end{APACrefauthors}%
\unskip\
\newblock
\APACrefYearMonthDay{2017}{}{}.
\newblock
{\BBOQ}\APACrefatitle {Giant ripples on comet 67P/Churyumov--Gerasimenko
  sculpted by sunset thermal wind} {Giant ripples on comet
  67p/churyumov--gerasimenko sculpted by sunset thermal wind}.{\BBCQ}
\newblock
\APACjournalVolNumPages{Proceedings of the National Academy of
  Sciences}{114}{10}{2509--2514}.
\PrintBackRefs{\CurrentBib}

\bibitem [\protect \citeauthoryear {%
Kennedy%
}{%
Kennedy%
}{%
{\protect \APACyear {1963}}%
}]{%
kennedy1963mechanics}
\APACinsertmetastar {%
kennedy1963mechanics}%
\begin{APACrefauthors}%
Kennedy, J\BPBI F.%
\end{APACrefauthors}%
\unskip\
\newblock
\APACrefYearMonthDay{1963}{}{}.
\newblock
{\BBOQ}\APACrefatitle {The mechanics of dunes and antidunes in erodible-bed
  channels} {The mechanics of dunes and antidunes in erodible-bed
  channels}.{\BBCQ}
\newblock
\APACjournalVolNumPages{Journal of Fluid mechanics}{16}{4}{521--544}.
\PrintBackRefs{\CurrentBib}

\bibitem [\protect \citeauthoryear {%
Kocurek%
, Ewing%
\BCBL {}\ \BBA {} Mohrig%
}{%
Kocurek%
\ \protect \BOthers {.}}{%
{\protect \APACyear {2010}}%
}]{%
kocurek2010bedform}
\APACinsertmetastar {%
kocurek2010bedform}%
\begin{APACrefauthors}%
Kocurek, G.%
, Ewing, R\BPBI C.%
\BCBL {}\ \BBA {} Mohrig, D.%
\end{APACrefauthors}%
\unskip\
\newblock
\APACrefYearMonthDay{2010}{}{}.
\newblock
{\BBOQ}\APACrefatitle {How do bedform patterns arise? New views on the role of
  bedform interactions within a set of boundary conditions} {How do bedform
  patterns arise? new views on the role of bedform interactions within a set of
  boundary conditions}.{\BBCQ}
\newblock
\APACjournalVolNumPages{Earth surface processes and landforms}{35}{1}{51--63}.
\PrintBackRefs{\CurrentBib}

\bibitem [\protect \citeauthoryear {%
Kouakou%
\ \BBA {} Lagr{\'e}e%
}{%
Kouakou%
\ \BBA {} Lagr{\'e}e%
}{%
{\protect \APACyear {2005}}%
}]{%
kouakou2005stability}
\APACinsertmetastar {%
kouakou2005stability}%
\begin{APACrefauthors}%
Kouakou, K\BPBI K\BPBI J.%
\BCBT {}\ \BBA {} Lagr{\'e}e, P\BHBI Y.%
\end{APACrefauthors}%
\unskip\
\newblock
\APACrefYearMonthDay{2005}{}{}.
\newblock
{\BBOQ}\APACrefatitle {Stability of an erodible bed in various shear flows}
  {Stability of an erodible bed in various shear flows}.{\BBCQ}
\newblock
\APACjournalVolNumPages{The European Physical Journal B-Condensed Matter and
  Complex Systems}{47}{1}{115--125}.
\PrintBackRefs{\CurrentBib}

\bibitem [\protect \citeauthoryear {%
Madole%
, Romig%
, Aleinikoff%
, VanSistine%
\BCBL {}\ \BBA {} Yacob%
}{%
Madole%
\ \protect \BOthers {.}}{%
{\protect \APACyear {2008}}%
}]{%
madole2008origin}
\APACinsertmetastar {%
madole2008origin}%
\begin{APACrefauthors}%
Madole, R\BPBI F.%
, Romig, J\BPBI H.%
, Aleinikoff, J\BPBI N.%
, VanSistine, D\BPBI P.%
\BCBL {}\ \BBA {} Yacob, E\BPBI Y.%
\end{APACrefauthors}%
\unskip\
\newblock
\APACrefYearMonthDay{2008}{}{}.
\newblock
{\BBOQ}\APACrefatitle {On the origin and age of the Great Sand Dunes, Colorado}
  {On the origin and age of the great sand dunes, colorado}.{\BBCQ}
\newblock
\APACjournalVolNumPages{Geomorphology}{99}{1-4}{99--119}.
\PrintBackRefs{\CurrentBib}

\bibitem [\protect \citeauthoryear {%
Naqshband%
, Hoitink%
, McElroy%
, Hurther%
\BCBL {}\ \BBA {} Hulscher%
}{%
Naqshband%
\ \protect \BOthers {.}}{%
{\protect \APACyear {2017}}%
}]{%
naqshband2017sharp}
\APACinsertmetastar {%
naqshband2017sharp}%
\begin{APACrefauthors}%
Naqshband, S.%
, Hoitink, A.%
, McElroy, B.%
, Hurther, D.%
\BCBL {}\ \BBA {} Hulscher, S\BPBI J.%
\end{APACrefauthors}%
\unskip\
\newblock
\APACrefYearMonthDay{2017}{}{}.
\newblock
{\BBOQ}\APACrefatitle {A sharp view on river dune transition to upper stage
  plane bed} {A sharp view on river dune transition to upper stage plane
  bed}.{\BBCQ}
\newblock
\APACjournalVolNumPages{Geophysical Research Letters}{44}{22}{11--437}.
\PrintBackRefs{\CurrentBib}

\bibitem [\protect \citeauthoryear {%
Narteau%
, Zhang%
, Rozier%
\BCBL {}\ \BBA {} Claudin%
}{%
Narteau%
\ \protect \BOthers {.}}{%
{\protect \APACyear {2009}}%
}]{%
narteau2009setting}
\APACinsertmetastar {%
narteau2009setting}%
\begin{APACrefauthors}%
Narteau, C.%
, Zhang, D.%
, Rozier, O.%
\BCBL {}\ \BBA {} Claudin, P.%
\end{APACrefauthors}%
\unskip\
\newblock
\APACrefYearMonthDay{2009}{}{}.
\newblock
{\BBOQ}\APACrefatitle {Setting the length and time scales of a cellular
  automaton dune model from the analysis of superimposed bed forms} {Setting
  the length and time scales of a cellular automaton dune model from the
  analysis of superimposed bed forms}.{\BBCQ}
\newblock
\APACjournalVolNumPages{Journal of Geophysical Research: Earth
  Surface}{114}{F3}{}.
\PrintBackRefs{\CurrentBib}

\bibitem [\protect \citeauthoryear {%
Nield%
}{%
Nield%
}{%
{\protect \APACyear {2011}}%
}]{%
nield2011surface}
\APACinsertmetastar {%
nield2011surface}%
\begin{APACrefauthors}%
Nield, J\BPBI M.%
\end{APACrefauthors}%
\unskip\
\newblock
\APACrefYearMonthDay{2011}{}{}.
\newblock
{\BBOQ}\APACrefatitle {Surface moisture- induced feedback in aeolian
  environments} {Surface moisture- induced feedback in aeolian
  environments}.{\BBCQ}
\newblock
\APACjournalVolNumPages{Geology}{39}{10}{915--918}.
\PrintBackRefs{\CurrentBib}

\bibitem [\protect \citeauthoryear {%
Nield%
\ \protect \BOthers {.}}{%
Nield%
\ \protect \BOthers {.}}{%
{\protect \APACyear {2013}}%
}]{%
nield2013estimating}
\APACinsertmetastar {%
nield2013estimating}%
\begin{APACrefauthors}%
Nield, J\BPBI M.%
, King, J.%
, Wiggs, G\BPBI F\BPBI S.%
, Leyland, J.%
, Bryant, R\BPBI G.%
, Chiverrell, R\BPBI C.%
\BDBL {}others%
\end{APACrefauthors}%
\unskip\
\newblock
\APACrefYearMonthDay{2013}{}{}.
\newblock
{\BBOQ}\APACrefatitle {Estimating aerodynamic roughness over complex surface
  terrain} {Estimating aerodynamic roughness over complex surface
  terrain}.{\BBCQ}
\newblock
\APACjournalVolNumPages{Journal of Geophysical Research:
  Atmospheres}{118}{23}{12--948}.
\PrintBackRefs{\CurrentBib}

\bibitem [\protect \citeauthoryear {%
Nield%
\ \BBA {} Wiggs%
}{%
Nield%
\ \BBA {} Wiggs%
}{%
{\protect \APACyear {2011}}%
}]{%
nield2011application}
\APACinsertmetastar {%
nield2011application}%
\begin{APACrefauthors}%
Nield, J\BPBI M.%
\BCBT {}\ \BBA {} Wiggs, G\BPBI F\BPBI S.%
\end{APACrefauthors}%
\unskip\
\newblock
\APACrefYearMonthDay{2011}{}{}.
\newblock
{\BBOQ}\APACrefatitle {The application of terrestrial laser scanning to aeolian
  saltation cloud measurement and its response to changing surface moisture}
  {The application of terrestrial laser scanning to aeolian saltation cloud
  measurement and its response to changing surface moisture}.{\BBCQ}
\newblock
\APACjournalVolNumPages{Earth Surface Processes and
  Landforms}{36}{2}{273--278}.
\PrintBackRefs{\CurrentBib}

\bibitem [\protect \citeauthoryear {%
P{\"a}htz%
\ \BBA {} Dur{\'a}n%
}{%
P{\"a}htz%
\ \BBA {} Dur{\'a}n%
}{%
{\protect \APACyear {2020}}%
}]{%
pahtz2020unification}
\APACinsertmetastar {%
pahtz2020unification}%
\begin{APACrefauthors}%
P{\"a}htz, T.%
\BCBT {}\ \BBA {} Dur{\'a}n, O.%
\end{APACrefauthors}%
\unskip\
\newblock
\APACrefYearMonthDay{2020}{}{}.
\newblock
{\BBOQ}\APACrefatitle {Unification of aeolian and fluvial sediment transport
  rate from granular physics} {Unification of aeolian and fluvial sediment
  transport rate from granular physics}.{\BBCQ}
\newblock
\APACjournalVolNumPages{Physical Review Letters}{124}{16}{168001}.
\PrintBackRefs{\CurrentBib}

\bibitem [\protect \citeauthoryear {%
P{\"a}htz%
, Kok%
, Parteli%
\BCBL {}\ \BBA {} Herrmann%
}{%
P{\"a}htz%
\ \protect \BOthers {.}}{%
{\protect \APACyear {2013}}%
}]{%
pahtz2013flux}
\APACinsertmetastar {%
pahtz2013flux}%
\begin{APACrefauthors}%
P{\"a}htz, T.%
, Kok, J\BPBI F.%
, Parteli, E\BPBI J.%
\BCBL {}\ \BBA {} Herrmann, H\BPBI J.%
\end{APACrefauthors}%
\unskip\
\newblock
\APACrefYearMonthDay{2013}{}{}.
\newblock
{\BBOQ}\APACrefatitle {Flux saturation length of sediment transport} {Flux
  saturation length of sediment transport}.{\BBCQ}
\newblock
\APACjournalVolNumPages{Physical review letters}{111}{21}{218002}.
\PrintBackRefs{\CurrentBib}

\bibitem [\protect \citeauthoryear {%
Parteli%
, Dur{\'a}n%
, Tsoar%
, Schw{\"a}mmle%
\BCBL {}\ \BBA {} Herrmann%
}{%
Parteli%
\ \protect \BOthers {.}}{%
{\protect \APACyear {2009}}%
}]{%
parteli2009dune}
\APACinsertmetastar {%
parteli2009dune}%
\begin{APACrefauthors}%
Parteli, E.%
, Dur{\'a}n, O.%
, Tsoar, H.%
, Schw{\"a}mmle, V.%
\BCBL {}\ \BBA {} Herrmann, H\BPBI J.%
\end{APACrefauthors}%
\unskip\
\newblock
\APACrefYearMonthDay{2009}{}{}.
\newblock
{\BBOQ}\APACrefatitle {Dune formation under bimodal winds} {Dune formation
  under bimodal winds}.{\BBCQ}
\newblock
\APACjournalVolNumPages{Proceedings of the National Academy of
  Sciences}{106}{52}{22085--22089}.
\PrintBackRefs{\CurrentBib}

\bibitem [\protect \citeauthoryear {%
Parteli%
, Schw{\"a}mmle%
, Herrmann%
, Monteiro%
\BCBL {}\ \BBA {} Maia%
}{%
Parteli%
\ \protect \BOthers {.}}{%
{\protect \APACyear {2006}}%
}]{%
parteli2006profile}
\APACinsertmetastar {%
parteli2006profile}%
\begin{APACrefauthors}%
Parteli, E.%
, Schw{\"a}mmle, V.%
, Herrmann, H\BPBI J.%
, Monteiro, L.%
\BCBL {}\ \BBA {} Maia, L.%
\end{APACrefauthors}%
\unskip\
\newblock
\APACrefYearMonthDay{2006}{}{}.
\newblock
{\BBOQ}\APACrefatitle {Profile measurement and simulation of a transverse dune
  field in the Len{\c{c}}{\'o}is Maranhenses} {Profile measurement and
  simulation of a transverse dune field in the len{\c{c}}{\'o}is
  maranhenses}.{\BBCQ}
\newblock
\APACjournalVolNumPages{Geomorphology}{81}{1-2}{29--42}.
\PrintBackRefs{\CurrentBib}

\bibitem [\protect \citeauthoryear {%
Pearce%
\ \BBA {} Walker%
}{%
Pearce%
\ \BBA {} Walker%
}{%
{\protect \APACyear {2005}}%
}]{%
pearce2005frequency}
\APACinsertmetastar {%
pearce2005frequency}%
\begin{APACrefauthors}%
Pearce, K\BPBI I.%
\BCBT {}\ \BBA {} Walker, I\BPBI J.%
\end{APACrefauthors}%
\unskip\
\newblock
\APACrefYearMonthDay{2005}{}{}.
\newblock
{\BBOQ}\APACrefatitle {Frequency and magnitude biases in the
  ‘Fryberger’model, with implications for characterizing geomorphically
  effective winds} {Frequency and magnitude biases in the ‘fryberger’model,
  with implications for characterizing geomorphically effective winds}.{\BBCQ}
\newblock
\APACjournalVolNumPages{Geomorphology}{68}{1-2}{39--55}.
\PrintBackRefs{\CurrentBib}

\bibitem [\protect \citeauthoryear {%
Phillips%
\ \protect \BOthers {.}}{%
Phillips%
\ \protect \BOthers {.}}{%
{\protect \APACyear {2019}}%
}]{%
phillips2019low}
\APACinsertmetastar {%
phillips2019low}%
\begin{APACrefauthors}%
Phillips, J.%
, Ewing, R.%
, Bowling, R.%
, Weymer, B\BPBI A.%
, Barrineau, P.%
, Nittrouer, J.%
\BCBL {}\ \BBA {} Everett, M.%
\end{APACrefauthors}%
\unskip\
\newblock
\APACrefYearMonthDay{2019}{}{}.
\newblock
{\BBOQ}\APACrefatitle {Low-angle eolian deposits formed by protodune migration,
  and insights into slipface development at White Sands Dune Field, New Mexico}
  {Low-angle eolian deposits formed by protodune migration, and insights into
  slipface development at white sands dune field, new mexico}.{\BBCQ}
\newblock
\APACjournalVolNumPages{Aeolian Research}{36}{}{9--26}.
\PrintBackRefs{\CurrentBib}

\bibitem [\protect \citeauthoryear {%
Ping%
, Narteau%
, Dong%
, Zhang%
\BCBL {}\ \BBA {} Courrech~du Pont%
}{%
Ping%
\ \protect \BOthers {.}}{%
{\protect \APACyear {2014}}%
}]{%
ping2014emergence}
\APACinsertmetastar {%
ping2014emergence}%
\begin{APACrefauthors}%
Ping, L.%
, Narteau, C.%
, Dong, Z.%
, Zhang, Z.%
\BCBL {}\ \BBA {} Courrech~du Pont, S.%
\end{APACrefauthors}%
\unskip\
\newblock
\APACrefYearMonthDay{2014}{}{}.
\newblock
{\BBOQ}\APACrefatitle {Emergence of oblique dunes in a landscape-scale
  experiment} {Emergence of oblique dunes in a landscape-scale
  experiment}.{\BBCQ}
\newblock
\APACjournalVolNumPages{Nature Geoscience}{7}{2}{99--103}.
\PrintBackRefs{\CurrentBib}

\bibitem [\protect \citeauthoryear {%
Ralaiarisoa%
\ \protect \BOthers {.}}{%
Ralaiarisoa%
\ \protect \BOthers {.}}{%
{\protect \APACyear {2020}}%
}]{%
ralaiarisoa2020transition}
\APACinsertmetastar {%
ralaiarisoa2020transition}%
\begin{APACrefauthors}%
Ralaiarisoa, J.%
, Besnard, J.%
, Furieri, B.%
, Dupont, P.%
, El~Moctar, A\BPBI O.%
, Naaim-Bouvet, F.%
\BCBL {}\ \BBA {} Valance, A.%
\end{APACrefauthors}%
\unskip\
\newblock
\APACrefYearMonthDay{2020}{}{}.
\newblock
{\BBOQ}\APACrefatitle {Transition from Saltation to Collisional Regime in
  Windblown Sand} {Transition from saltation to collisional regime in windblown
  sand}.{\BBCQ}
\newblock
\APACjournalVolNumPages{Physical Review Letters}{124}{19}{198501}.
\PrintBackRefs{\CurrentBib}

\bibitem [\protect \citeauthoryear {%
Rasmussen%
, Iversen%
\BCBL {}\ \BBA {} Rautahemio%
}{%
Rasmussen%
\ \protect \BOthers {.}}{%
{\protect \APACyear {1996}}%
}]{%
rasmussen1996saltation}
\APACinsertmetastar {%
rasmussen1996saltation}%
\begin{APACrefauthors}%
Rasmussen, K\BPBI R.%
, Iversen, J\BPBI D.%
\BCBL {}\ \BBA {} Rautahemio, P.%
\end{APACrefauthors}%
\unskip\
\newblock
\APACrefYearMonthDay{1996}{}{}.
\newblock
{\BBOQ}\APACrefatitle {Saltation and wind-flow interaction in a variable slope
  wind tunnel} {Saltation and wind-flow interaction in a variable slope wind
  tunnel}.{\BBCQ}
\newblock
\APACjournalVolNumPages{Geomorphology}{17}{1-3}{19--28}.
\PrintBackRefs{\CurrentBib}

\bibitem [\protect \citeauthoryear {%
Reffet%
, Courrech~du Pont%
, Hersen%
\BCBL {}\ \BBA {} Douady%
}{%
Reffet%
\ \protect \BOthers {.}}{%
{\protect \APACyear {2010}}%
}]{%
reffet2010formation}
\APACinsertmetastar {%
reffet2010formation}%
\begin{APACrefauthors}%
Reffet, E.%
, Courrech~du Pont, S.%
, Hersen, P.%
\BCBL {}\ \BBA {} Douady, S.%
\end{APACrefauthors}%
\unskip\
\newblock
\APACrefYearMonthDay{2010}{}{}.
\newblock
{\BBOQ}\APACrefatitle {Formation and stability of transverse and longitudinal
  sand dunes} {Formation and stability of transverse and longitudinal sand
  dunes}.{\BBCQ}
\newblock
\APACjournalVolNumPages{Geology}{38}{6}{491--494}.
\PrintBackRefs{\CurrentBib}

\bibitem [\protect \citeauthoryear {%
Richards%
}{%
Richards%
}{%
{\protect \APACyear {1980}}%
}]{%
richards1980formation}
\APACinsertmetastar {%
richards1980formation}%
\begin{APACrefauthors}%
Richards, K\BPBI J.%
\end{APACrefauthors}%
\unskip\
\newblock
\APACrefYearMonthDay{1980}{}{}.
\newblock
{\BBOQ}\APACrefatitle {The formation of ripples and dunes on an erodible bed}
  {The formation of ripples and dunes on an erodible bed}.{\BBCQ}
\newblock
\APACjournalVolNumPages{Journal of Fluid Mechanics}{99}{3}{597--618}.
\PrintBackRefs{\CurrentBib}

\bibitem [\protect \citeauthoryear {%
Rozier%
\ \BBA {} Narteau%
}{%
Rozier%
\ \BBA {} Narteau%
}{%
{\protect \APACyear {2014}}%
}]{%
rozier2014real}
\APACinsertmetastar {%
rozier2014real}%
\begin{APACrefauthors}%
Rozier, O.%
\BCBT {}\ \BBA {} Narteau, C.%
\end{APACrefauthors}%
\unskip\
\newblock
\APACrefYearMonthDay{2014}{}{}.
\newblock
{\BBOQ}\APACrefatitle {A real-space cellular automaton laboratory} {A
  real-space cellular automaton laboratory}.{\BBCQ}
\newblock
\APACjournalVolNumPages{Earth Surface Processes and Landforms}{39}{1}{98--109}.
\PrintBackRefs{\CurrentBib}

\bibitem [\protect \citeauthoryear {%
Rubin%
\ \BBA {} Hunter%
}{%
Rubin%
\ \BBA {} Hunter%
}{%
{\protect \APACyear {1987}}%
}]{%
rubin1987bedform}
\APACinsertmetastar {%
rubin1987bedform}%
\begin{APACrefauthors}%
Rubin, D\BPBI M.%
\BCBT {}\ \BBA {} Hunter, R\BPBI E.%
\end{APACrefauthors}%
\unskip\
\newblock
\APACrefYearMonthDay{1987}{}{}.
\newblock
{\BBOQ}\APACrefatitle {Bedform alignment in directionally varying flows}
  {Bedform alignment in directionally varying flows}.{\BBCQ}
\newblock
\APACjournalVolNumPages{Science}{237}{4812}{276--278}.
\PrintBackRefs{\CurrentBib}

\bibitem [\protect \citeauthoryear {%
Rubin%
\ \BBA {} Ikeda%
}{%
Rubin%
\ \BBA {} Ikeda%
}{%
{\protect \APACyear {1990}}%
}]{%
rubin1990flume}
\APACinsertmetastar {%
rubin1990flume}%
\begin{APACrefauthors}%
Rubin, D\BPBI M.%
\BCBT {}\ \BBA {} Ikeda, H.%
\end{APACrefauthors}%
\unskip\
\newblock
\APACrefYearMonthDay{1990}{}{}.
\newblock
{\BBOQ}\APACrefatitle {Flume experiments on the alignment of transverse,
  oblique, and longitudinal dunes in directionally varying flows} {Flume
  experiments on the alignment of transverse, oblique, and longitudinal dunes
  in directionally varying flows}.{\BBCQ}
\newblock
\APACjournalVolNumPages{Sedimentology}{37}{4}{673--684}.
\PrintBackRefs{\CurrentBib}

\bibitem [\protect \citeauthoryear {%
Sauermann%
, Kroy%
\BCBL {}\ \BBA {} Herrmann%
}{%
Sauermann%
\ \protect \BOthers {.}}{%
{\protect \APACyear {2001}}%
}]{%
sauermann2001continuum}
\APACinsertmetastar {%
sauermann2001continuum}%
\begin{APACrefauthors}%
Sauermann, G.%
, Kroy, K.%
\BCBL {}\ \BBA {} Herrmann, H\BPBI J.%
\end{APACrefauthors}%
\unskip\
\newblock
\APACrefYearMonthDay{2001}{}{}.
\newblock
{\BBOQ}\APACrefatitle {Continuum saltation model for sand dunes} {Continuum
  saltation model for sand dunes}.{\BBCQ}
\newblock
\APACjournalVolNumPages{Physical Review E}{64}{3}{031305}.
\PrintBackRefs{\CurrentBib}

\bibitem [\protect \citeauthoryear {%
Selmani%
, Valance%
, Ould El~Moctar%
, Dupont%
\BCBL {}\ \BBA {} Zegadi%
}{%
Selmani%
\ \protect \BOthers {.}}{%
{\protect \APACyear {2018}}%
}]{%
selmani2018aeolian}
\APACinsertmetastar {%
selmani2018aeolian}%
\begin{APACrefauthors}%
Selmani, H.%
, Valance, A.%
, Ould El~Moctar, A.%
, Dupont, P.%
\BCBL {}\ \BBA {} Zegadi, R.%
\end{APACrefauthors}%
\unskip\
\newblock
\APACrefYearMonthDay{2018}{}{}.
\newblock
{\BBOQ}\APACrefatitle {Aeolian sand transport in out-of-equilibrium regimes}
  {Aeolian sand transport in out-of-equilibrium regimes}.{\BBCQ}
\newblock
\APACjournalVolNumPages{Geophysical Research Letters}{45}{4}{1838--1844}.
\PrintBackRefs{\CurrentBib}

\bibitem [\protect \citeauthoryear {%
Stockton%
\ \BBA {} Gillette%
}{%
Stockton%
\ \BBA {} Gillette%
}{%
{\protect \APACyear {1990}}%
}]{%
stockton1990field}
\APACinsertmetastar {%
stockton1990field}%
\begin{APACrefauthors}%
Stockton, P.%
\BCBT {}\ \BBA {} Gillette, D.%
\end{APACrefauthors}%
\unskip\
\newblock
\APACrefYearMonthDay{1990}{}{}.
\newblock
{\BBOQ}\APACrefatitle {Field measurement of the sheltering effect of vegetation
  on erodible land surfaces} {Field measurement of the sheltering effect of
  vegetation on erodible land surfaces}.{\BBCQ}
\newblock
\APACjournalVolNumPages{Land Degradation \& Development}{2}{2}{77--85}.
\PrintBackRefs{\CurrentBib}

\bibitem [\protect \citeauthoryear {%
Tsoar%
}{%
Tsoar%
}{%
{\protect \APACyear {2005}}%
}]{%
tsoar2005sand}
\APACinsertmetastar {%
tsoar2005sand}%
\begin{APACrefauthors}%
Tsoar, H.%
\end{APACrefauthors}%
\unskip\
\newblock
\APACrefYearMonthDay{2005}{}{}.
\newblock
{\BBOQ}\APACrefatitle {Sand dunes mobility and stability in relation to
  climate} {Sand dunes mobility and stability in relation to climate}.{\BBCQ}
\newblock
\APACjournalVolNumPages{Physica A: Statistical Mechanics and its
  Applications}{357}{1}{50--56}.
\PrintBackRefs{\CurrentBib}

\bibitem [\protect \citeauthoryear {%
Ungar%
\ \BBA {} Haff%
}{%
Ungar%
\ \BBA {} Haff%
}{%
{\protect \APACyear {1987}}%
}]{%
ungar1987steady}
\APACinsertmetastar {%
ungar1987steady}%
\begin{APACrefauthors}%
Ungar, J\BPBI E.%
\BCBT {}\ \BBA {} Haff, P.%
\end{APACrefauthors}%
\unskip\
\newblock
\APACrefYearMonthDay{1987}{}{}.
\newblock
{\BBOQ}\APACrefatitle {Steady state saltation in air} {Steady state saltation
  in air}.{\BBCQ}
\newblock
\APACjournalVolNumPages{Sedimentology}{34}{2}{289--299}.
\PrintBackRefs{\CurrentBib}

\bibitem [\protect \citeauthoryear {%
Valdez%
}{%
Valdez%
}{%
{\protect \APACyear {1996}}%
}]{%
valdez1996role}
\APACinsertmetastar {%
valdez1996role}%
\begin{APACrefauthors}%
Valdez, A\BPBI D.%
\end{APACrefauthors}%
\unskip\
\newblock
\APACrefYearMonthDay{1996}{}{}.
\newblock
{\BBOQ}\APACrefatitle {The role of streams in the development of the Great Sand
  Dunes and their connection with the hydrologic cycle} {The role of streams in
  the development of the great sand dunes and their connection with the
  hydrologic cycle}.{\BBCQ}
\newblock
\APACjournalVolNumPages{Great Sand Dunes National Monument, Mosca, Colorado.
  Available: http://www. nps. gov/grsa/naturescience/upload/Trp2029.
  pdf}{}{}{}.
\PrintBackRefs{\CurrentBib}

\bibitem [\protect \citeauthoryear {%
Walker%
\ \BBA {} Nickling%
}{%
Walker%
\ \BBA {} Nickling%
}{%
{\protect \APACyear {2002}}%
}]{%
walker2002dynamics}
\APACinsertmetastar {%
walker2002dynamics}%
\begin{APACrefauthors}%
Walker, I\BPBI J.%
\BCBT {}\ \BBA {} Nickling, W\BPBI G.%
\end{APACrefauthors}%
\unskip\
\newblock
\APACrefYearMonthDay{2002}{}{}.
\newblock
{\BBOQ}\APACrefatitle {Dynamics of secondary airflow and sediment transport
  over and in the lee of transverse dunes} {Dynamics of secondary airflow and
  sediment transport over and in the lee of transverse dunes}.{\BBCQ}
\newblock
\APACjournalVolNumPages{Progress in Physical Geography}{26}{1}{47--75}.
\PrintBackRefs{\CurrentBib}

\bibitem [\protect \citeauthoryear {%
Werner%
\ \BBA {} Kocurek%
}{%
Werner%
\ \BBA {} Kocurek%
}{%
{\protect \APACyear {1999}}%
}]{%
werner1999bedform}
\APACinsertmetastar {%
werner1999bedform}%
\begin{APACrefauthors}%
Werner, B.%
\BCBT {}\ \BBA {} Kocurek, G.%
\end{APACrefauthors}%
\unskip\
\newblock
\APACrefYearMonthDay{1999}{}{}.
\newblock
{\BBOQ}\APACrefatitle {Bedform spacing from defect dynamics} {Bedform spacing
  from defect dynamics}.{\BBCQ}
\newblock
\APACjournalVolNumPages{Geology}{27}{8}{727--730}.
\PrintBackRefs{\CurrentBib}

\bibitem [\protect \citeauthoryear {%
Wiggs%
, Livingstone%
\BCBL {}\ \BBA {} Warren%
}{%
Wiggs%
\ \protect \BOthers {.}}{%
{\protect \APACyear {1996}}%
}]{%
wiggs1996role}
\APACinsertmetastar {%
wiggs1996role}%
\begin{APACrefauthors}%
Wiggs, G\BPBI F\BPBI S.%
, Livingstone, I.%
\BCBL {}\ \BBA {} Warren, A.%
\end{APACrefauthors}%
\unskip\
\newblock
\APACrefYearMonthDay{1996}{}{}.
\newblock
{\BBOQ}\APACrefatitle {The role of streamline curvature in sand dune dynamics:
  evidence from field and wind tunnel measurements} {The role of streamline
  curvature in sand dune dynamics: evidence from field and wind tunnel
  measurements}.{\BBCQ}
\newblock
\APACjournalVolNumPages{Geomorphology}{17}{1-3}{29--46}.
\PrintBackRefs{\CurrentBib}

\bibitem [\protect \citeauthoryear {%
Wiggs%
\ \BBA {} Weaver%
}{%
Wiggs%
\ \BBA {} Weaver%
}{%
{\protect \APACyear {2012}}%
}]{%
wiggs2012turbulent}
\APACinsertmetastar {%
wiggs2012turbulent}%
\begin{APACrefauthors}%
Wiggs, G\BPBI F\BPBI S.%
\BCBT {}\ \BBA {} Weaver, C\BPBI M.%
\end{APACrefauthors}%
\unskip\
\newblock
\APACrefYearMonthDay{2012}{}{}.
\newblock
{\BBOQ}\APACrefatitle {Turbulent flow structures and aeolian sediment transport
  over a barchan sand dune} {Turbulent flow structures and aeolian sediment
  transport over a barchan sand dune}.{\BBCQ}
\newblock
\APACjournalVolNumPages{Geophysical research letters}{39}{5}{}.
\PrintBackRefs{\CurrentBib}

\bibitem [\protect \citeauthoryear {%
Wilcock%
, Orr%
\BCBL {}\ \BBA {} Marr%
}{%
Wilcock%
\ \protect \BOthers {.}}{%
{\protect \APACyear {2008}}%
}]{%
wilcock2008need}
\APACinsertmetastar {%
wilcock2008need}%
\begin{APACrefauthors}%
Wilcock, P\BPBI R.%
, Orr, C\BPBI H.%
\BCBL {}\ \BBA {} Marr, J\BPBI D.%
\end{APACrefauthors}%
\unskip\
\newblock
\APACrefYearMonthDay{2008}{}{}.
\newblock
{\BBOQ}\APACrefatitle {The need for full-scale experiments in river science}
  {The need for full-scale experiments in river science}.{\BBCQ}
\newblock
\APACjournalVolNumPages{Eos, Transactions American Geophysical
  Union}{89}{1}{6--6}.
\PrintBackRefs{\CurrentBib}

\bibitem [\protect \citeauthoryear {%
Worman%
, Murray%
, Littlewood%
, Andreotti%
\BCBL {}\ \BBA {} Claudin%
}{%
Worman%
\ \protect \BOthers {.}}{%
{\protect \APACyear {2013}}%
}]{%
worman2013modeling}
\APACinsertmetastar {%
worman2013modeling}%
\begin{APACrefauthors}%
Worman, S\BPBI L.%
, Murray, A\BPBI B.%
, Littlewood, R.%
, Andreotti, B.%
\BCBL {}\ \BBA {} Claudin, P.%
\end{APACrefauthors}%
\unskip\
\newblock
\APACrefYearMonthDay{2013}{}{}.
\newblock
{\BBOQ}\APACrefatitle {Modeling emergent large-scale structures of barchan dune
  fields} {Modeling emergent large-scale structures of barchan dune
  fields}.{\BBCQ}
\newblock
\APACjournalVolNumPages{Geology}{41}{10}{1059--1062}.
\PrintBackRefs{\CurrentBib}

\bibitem [\protect \citeauthoryear {%
Zhang%
, Narteau%
\BCBL {}\ \BBA {} Rozier%
}{%
Zhang%
\ \protect \BOthers {.}}{%
{\protect \APACyear {2010}}%
}]{%
zhang2010morphodynamics}
\APACinsertmetastar {%
zhang2010morphodynamics}%
\begin{APACrefauthors}%
Zhang, D.%
, Narteau, C.%
\BCBL {}\ \BBA {} Rozier, O.%
\end{APACrefauthors}%
\unskip\
\newblock
\APACrefYearMonthDay{2010}{}{}.
\newblock
{\BBOQ}\APACrefatitle {Morphodynamics of barchan and transverse dunes using a
  cellular automaton model} {Morphodynamics of barchan and transverse dunes
  using a cellular automaton model}.{\BBCQ}
\newblock
\APACjournalVolNumPages{Journal of Geophysical Research: Earth
  Surface}{115}{F3}{}.
\PrintBackRefs{\CurrentBib}

\end{thebibliography}

\clearpage

\section*{Supporting Information for "Dune initiation in a bimodal wind regime"}

\noindent\textbf{Contents of this file}
\begin{enumerate}
\item Text S1 to S2
\item Figure S1  

\end{enumerate}

\noindent\textbf{Introduction}

\begin{figure}[h]
\centering
\includegraphics[width=0.9\linewidth]{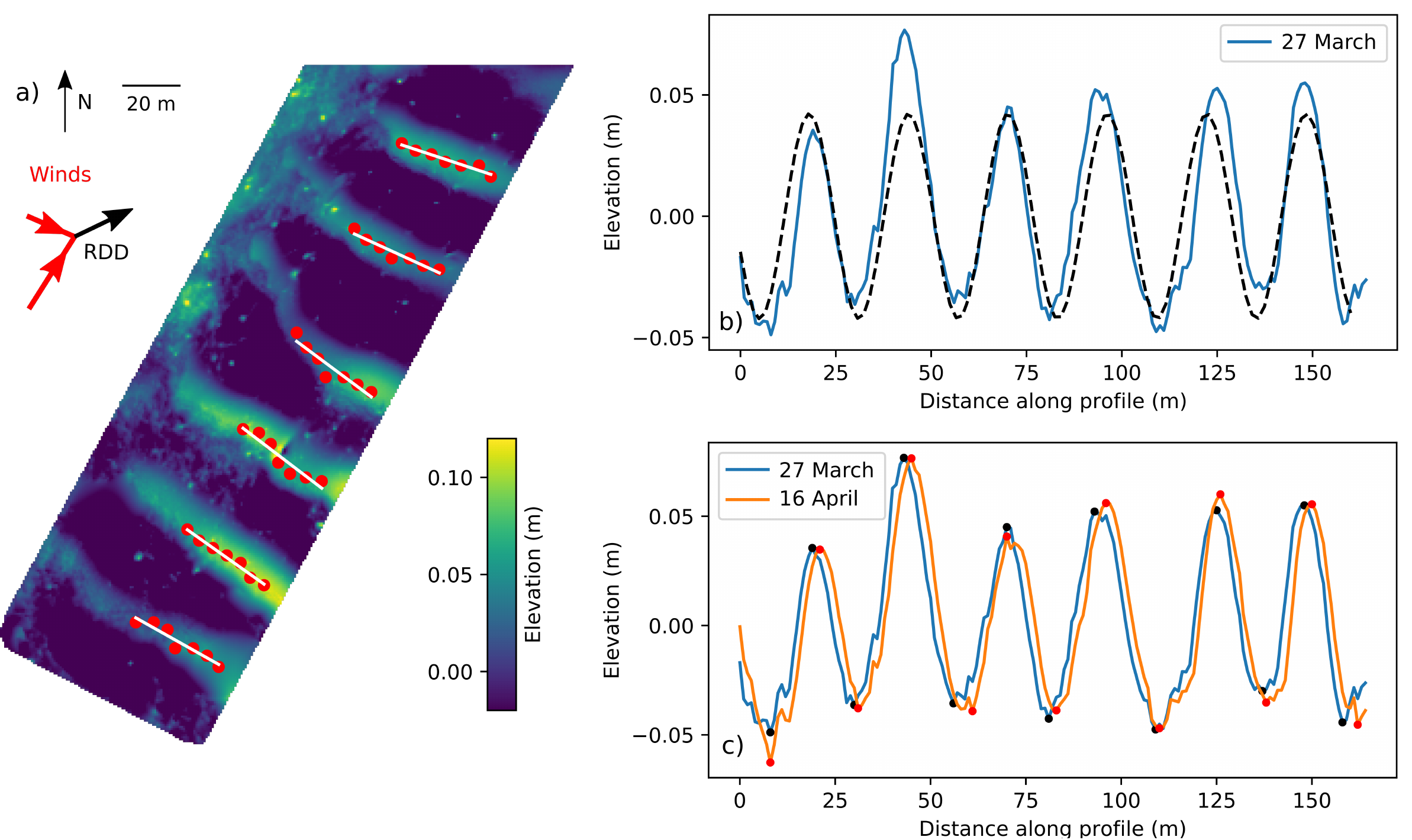}
\caption{a) DEM of the field area on 27 March 2019, the red dots represent the peak position and the white lines the linear fit used to determine the crest orientation. b) Bed elevations measured on 27 March 2019, with the sinusoidal fit (black dotted line). c) Bed elevations measured on 27 March 2019 and 16 April 2019, the black and red dots are the peak positions (black: 27 March 2019 and red: 16 April 2019). \label{fig:supp}}
\end{figure}

In this supporting information, we provide details on the linear stability analysis for a uni-directional wind (Text S1) and on the measurements of the morphological characteristics (Text S2). Figure \ref{fig:supp} provides example of the morphological measurements.

%

\noindent\textbf{Text S1. Linear stability analysis \label{S1}}

For a uni-directional wind following \citeA{andreotti2012bedforms}, the dispersion relation can be expressed in a dimensionless form as,

\begin{equation}
\begin{array}{l}
\sigma = \frac{k^2}{1 + (k\,\mathrm{cos}\, \alpha)^2}\left[ \mathrm{cos}\,\alpha({\cal{B}} - k\,\mathrm{cos}\,\alpha\, {{\cal{A}}} ) \left( \mathrm{cos}^2\alpha + \frac{r}{2} \mathrm{sin}^2\alpha\right)\right.\\\\
\left. - \frac{1}{\mu} \frac{u_{*\mathrm{th}}}{u_*} \left(\frac{u_{*\mathrm{th}}}{u_*} \mathrm{cos}^2\alpha + r\, \mathrm{sin}^2\alpha\right) \right], 
\end{array}
\label{eq:growth_rate_uni}    
\end{equation}

and, 

\begin{equation}
\begin{array}{l}
c = \frac{k}{1 + (k\, \mathrm{cos}\, \alpha)^2}\left[ \mathrm{cos}\,\alpha({\cal{A}}- k\,\mathrm{cos}\,\alpha\, {{\cal{B}}} ) \left( \mathrm{cos}^2\alpha + \frac{r}{2} \mathrm{sin}^2\alpha\right)\right.\\\\
\left. - k\, \mathrm{cos}\, \alpha\, \frac{1}{\mu} \frac{u_{*\mathrm{th}}}{u_*} \left(\frac{u_{*\mathrm{th}}}{u_*} \mathrm{cos}^2\alpha + r\, \mathrm{sin}^2\alpha\right) \right], 
\end{array}
\label{eq:propagation_velocity_uni}
\end{equation}

where $k$ is the wavenumber, such as the wavelength $\lambda = 2 \pi / k$, $\alpha$ is the angle between the pattern and the perpendicular to the resultant flux direction, ${\cal{A}}$ and ${\cal{B}}$ are the hydrodynamic coefficients of the linear stability when the dune crests are perpendicular to the flow, and $\mu$ is the avalanche slope (0.7). In these equations, lengths are rescaled  by $L_{\mathrm{sat}}$ and times by $L^2_{\mathrm{sat}}(u^2_* - u^2_{*\mathrm{th}})/(Q_s u^2_*)$. 

To obtain an agreement between morphological measurement and calculation, we selected values of $\cal{A}$ and $\cal{B}$ ranging from [3.2, 4] and [1.6, 2.1] respectively, which corresponds to a turbulent regime  according to \citeA{charru2013sand}. The coefficients $\cal{A}$ and $\cal{B}$ we used in these study are of the same order as the ones calculated in the field by \citeA{claudin2013field} and estimated using the linear stability analysis by \citeA{gadalspatial}.

\noindent\textbf{Text S2. Measurement of the morphological characteristics \label{S2}} 

\textbf{Crest orientation:} To determine the crest orientation, we used a linear fit through the position of the crest for each protodune on each DEM, the slope of the linear fit gives us a direct measurement of the crest orientation (Fig. \ref{fig:supp}a).

\textbf{Dune wavelength:} The dune wavelength is measured by using the difference between each peak position on each DEM. The value calculated quantitatively agrees with the one given by the sinusoidal fit, $\lambda_{fit}$ = 26.06 m (Fig. \ref{fig:supp}b).

\textbf{Growth rate: } Following \citeA{gao2015phase}, the dune growth rate is defined such as,
\begin{equation}
\sigma = \frac{1}{H}\frac{\Delta H}{\Delta t},
\end{equation}
with $H$ the dune elevation on the first DEM (27 March 2019), $\Delta H$ the difference in dune elevation between the first and the last DEM (27 March 2019 and 16 April 2019), and $\Delta t$ the time between the first and the last DEM (19 days). We used the difference between the maximum and the minimum peak position for each protodune to determine their height, $H$ (Fig. \ref{fig:supp}c).

\textbf{Migration rate:} To determine the migration rate we calculated the difference in the peak postion for each protodune between the first and the last DEM (Fig. \ref{fig:supp}c). The value calculed using the difference between peak position quantitatively agrees with the one calculated using the phase difference between the two sinusoidal fits, $c_{fit}$ = 0.1 m$\,$day$^{-1}$ (Fig. \ref{fig:supp}b).

\end{document}